\documentclass{article}
\usepackage{amsfonts}
\usepackage{amsmath}
\usepackage{hyperref}
\usepackage{curves}
\usepackage{enumerate}
\usepackage[utf8]{inputenc}
\usepackage{latexsym}
\usepackage{braket}
\usepackage{mathrsfs}
\usepackage{mathtools}
\usepackage{color}
\usepackage{color,soul}
\usepackage{xcolor}
\usepackage{graphicx}
\usepackage{subcaption}
\usepackage{enumitem}
\usepackage[toc,page]{appendix}
\usepackage[a4paper, total={6in, 8in}]{geometry}

\usepackage{cite}
\usepackage[english]{babel}
\usepackage{babelbib}
\usepackage{multirow}
\usepackage{tabularx}
\usepackage{makecell}
\setcellgapes{4pt}
\usepackage[thinlines]{easytable}

\usepackage{colortbl}
\def\bal#1\eal{\begin{align}#1\end{align}}
\usepackage{color,soul}
\newcommand{\bsub}{\begin{subequations}}
\newcommand{\esub}{\end{subequations}}
\def\bal#1\eal{\begin{align}#1\end{align}}

\newcommand{\gra}{{\alpha}} \newcommand{\grb}{{\beta}} \newcommand{\grg}{{\gamma}} \newcommand{\grd}{{\delta}}
 \newcommand{\grz}{{\zeta}}  \newcommand{\gru}{{\theta}}
  \newcommand{\grl}{{\lambda}} \newcommand{\grm}{{\mu}}
\newcommand{\grn}{{\nu}} \newcommand{\grj}{{\xi}}  \newcommand{\grp}{{\pi}}
\newcommand{\grr}{{\rho}} \newcommand{\grs}{{\sigma}} \newcommand{\grt}{{\tau}} \newcommand{\gry}{{\upsilon}}
\newcommand{\grf}{{\phi}}  \newcommand{\grc}{{\psi}} \newcommand{\grv}{{\omega}}

  \newcommand{\grL}{{\Lambda}}

\newcommand{\grF}{{\Phi}}   \newcommand{\grV}{{\Omega}}
\newcommand{\barr}{\overline}
\begin{document}

\title{\textbf{Classical and quantum analysis of 3D electromagnetic pp-wave spacetime}}
\author{ \textbf{T. Pailas}$^a$\thanks{teopailas879@hotmail.com} ,\,\,\textbf{N. Dimakis}$^b$\thanks{nsdimakis@scu.edu.cn},\,\,\textbf{A. Karagiorgos}$^a$\thanks{alexkarag@phys.uoa.gr},\,\,\textbf{Petros A. Terzis}$^a$\thanks{pterzis@phys.uoa.gr},\\\textbf{G.O. Papadopoulos}$^a$\thanks{gop@disroot.org},\,\,\textbf{T. Christodoulakis}$^a$\thanks{tchris@phys.uoa.gr}\\
\normalsize
{$^a$\it Department of Nuclear and Particle Physics, Faculty of Physics,}\\
\normalsize{\it National and Kapodistrian University of Athens, Athens 15784, Greece}\\
\normalsize
{$^b$\it Center for Theoretical Physics, College of Physical Science and Technology}\\
\normalsize{\it Sichuan University, Chengdu 610065, China}}

\maketitle

\abstract{The general classical solution of the 3D electromagnetic pp-wave spacetime has been obtained. The relevant line element contains an arbitrary essential function providing an infinite number of in-equivalent  geometries as solutions. A classification is presented based on the symmetry group. To the best of our knowledge, the solution corresponding to only one of the Classes is known. The dynamics of some of the Classes was also derived from a minisuperspace Lagrangian which has been constructed. This Lagrangian contains a degree of freedom (the lapse) which can be considered either as dynamical or non-dynamical (indicating a singular or a regular Lagrangian correspondingly). Surprisingly enough, on the space of classical solutions, an equivalence of these two points of view can be established. The canonical quantization is then used in order to quantize the system for both the singular and regular Hamiltonian. A subsequent interpretation of quantum states is based on a Bohm-like analysis. The semi-classical trajectories deviate from the classical only for the regular Hamiltonian and in particular for a superposition of eigenstates (a Gaussian initial state has been used). Thus, the above mentioned equivalence is broken at the quantum level. It is noteworthy that the semi-classical trajectories tend to the classical ones in the limit where the initial wavepacket is widely spread. Hence, even with this simple superposition state, the classical solutions are acquired as a limit of the semi-classical.}

\newpage

\section{Introduction}

The concept of gravitational waves in the weak field approximation, was predicted theoretically by Einstein himself, since the early days of General Relativity \cite{1916SPAW.......688E},\cite{1918SPAW.......154E}. A criticism on Einstein's gravitational wave solutions as well as the speed that those waves propagate was presented by Eddington in \cite{1922RSPSA.102..268E}. Later on, Einstein and Rosen published a paper predicting the existence of cylindrical gravitational waves, firstly denoted by them as a non physical solution due to the presence of a singularity \cite{EINSTEIN193743}. As it was pointed out by  Howard P. Robertson with a letter to the authors, this was a mere coordinate singularity related to the cylindrical coordinates. Since then, a lot of work has been done in this field, with the highlights being the indirect detection of gravitational waves by Russel Alan Hulse and Joseph Hooton Taylor Jr. through the energy and angular momentum loss of a binary pulsar \cite{Taylor:1979zz},\cite{Taylor:1982zz}, while the first direct came in 2015 from the merger of two black holes \cite{PhysRevLett.116.061102}.

All those works and even more mentioned in the previous paragraph, were dealing with the linearized Einstein's field equations and every solution found was valid in the weak field limit. The existence of exact analytic solutions with wave characteristics has also being proven theoretically. The first to appear was due to Bringmann \cite{Brinkmann1925} were he studied conformal maps between Einstein spaces, and Baldwin and Jeffery in \cite{1926RSPSA.111...95B} were they coupled gravity and electromagnetism with the ansatz of plane wave propagating at some specific velocity. Due to difficulties on defining the notion of a wave in the absence of a fixed background, a geometrical definition was needed. Petrov was able to algebraically classify the Weyl tensor, which is the part of the Riemann tensor which is not affected by the presence of the local matter \cite{2000GReGr..32.1665P}. In his notation, the type N was associated with gravitational radiation in a way similar to that of electromagnetic waves, which are characterized by ``null'' Maxwell fields. Later on, Sachs introduced the gravitational ray optics \cite{Sachs309} and Kundt studied plane-fronted gravitational waves \cite{Kundt1961}. A special class of those are called pp-waves which have the property of all their first kind curvature scalars being equal to zero. Another interesting fact concerning those spacetimes is that they are not globally hyperbolic as was pointed out by Roger Penrose \cite{RevModPhys.37.215}. The pp-waves appear in a quite broad area, from general relativity to string and brane theory as well as supergravity, so is nearly impossible to cite each and every work. We choose only few representatives. The authors of \cite{PhysRevD.43.3907} have cast Newton's equations for N non-relativistic interacting particles, according to the inverse square law, in the form of equations for null geodesics in a $(3N+2)$-dimensional generalized pp-wave spacetime. In \cite{DIAMANDIS199269} the authors have proven that the conformal invariance of a $\grs$- model does not imply only a Robertson-Walker universe but also a special kind of a pp-wave metric. For a bosonic string in a specially chosen pp-wave spacetime the conformal anomaly vanishes as it was shown in \cite{Duval:1993pa}, while pp-waves are shown to be suitable, up to conditions, for providing string vacua for the sigma model perturbation expansion, at all orders \cite{Duval:1993tg}. Brinkmann spacetimes (pp-waves) whose extra components of the metric can be viewed as external gauge fields with the mass as coupling constant where investigated in\cite{Duval:1994qye}. The Penrose Limits of M-theory backgrounds were investigated in \cite{0264-9381-19-18-310}. Brandhuber and Sfetsos studied pp-waves which arose as Penrose limits of D3-branes \cite{Brandhuber:2002wd}. When it comes to the geodesic deviation equation, Svarc and Podolsky studied it in Kundt spacetimes of any dimension as is indicated in their work \cite{Svarc:2012se}. The analog of pp-waves was introduced in Finsler geometry by Fuster and Pabst in \cite{Fuster:2015tua}. More recently, the authors of \cite{Paliathanasis:2017kzv} classify the symmetries of Wave and Klein-Gordon equations in the pp-wave spacetimes. The geometry of a pp-wave, appeared as the configuration space of the minisuperspace Lagrangian of a Bianchi III LRS geometry coupled to electromagnetic field, as it was found in \cite{Karagiorgos:2017nta}. Finally, some of the solutions of four and five dimensional vacuum Bianchi types were identified as pp-waves in the following publications \cite{Christodoulakis:2004yx},\cite{Pailas:2018tzy}.

So far, it is seems like our world is best described by four dimensions. There are no experimental evidence to support something else. Nevertheless, studying the laws of physics in various dimensions provide us with insights. For instance, the study of $(2+1)$-dimensional gravity is of interest due to the absence of gravitational local degrees of freedom. Edward Witten studied it's relation with Chern-Simons theory under a specific gauge group \cite{1988NuPhB.311...46W},\cite{Witten:2007kt}. The most well known solution is the BTZ black hole named after the authors Maximo Ba\~nados, Claudio Teitelboim and Jorge Zanelli \cite{Banados:1992wn}. In \cite{Deser:2004wd}, the authors studied the conformal gravity in $(2+1)$ dimensions, where the Cotton tensor was equated to the energy momentum tensor of an improved scalar field. Exact solutions were found in the pp-wave regime and equivalence with topological massive gravity was achieved. A variety of solutions concerning with $(2+1)$ gravity can be found in the following book \cite{Garcia-Diaz:2017cpv}. Finally, a more recent publication provides an insight to the solution space in $2+1$ dimensions for $\grL$-vacuum, pure radiation and gyratons \cite{Podolsky:2018zha}.

In this particular work, we intend to study the (2+1)-dimensional pp-wave spacetimes in the presence of electromagnetic field, both at the classical and quantum regime. For the classical solutions to be found, we would like to provide a classification based on the number of the Killing fields and their Lie algebra. Furthermore, we wish to construct a minisuperspace Lagrangian and it's corresponding Hamiltonian. With this at hand, we will canonically quantize the system. For the quantum states, the semi-classical analysis will be used in order to interpret the solutions in geometrical terms.

The paper's structure is organized as follows. There are five main sections, each one contains a number of subsections. In section $2$ the classical description of the system is provided. The general solution is found and classified based on the number and the form of the Lie algebra of the existing Killing fields. For some of the Classes appeared in the classification, a minisuperspace Lagrangian and Hamiltonian is provided. The quantum analysis is presented in section $3$. Specifically, the eigenfunctions have been found for each of the sub-algebras; the sub-algebras consist of the commuting operators corresponding to the Hamiltonian and the conserved Noether charges. A possible interpretation of the quantum solutions is given in section $4$ in the context of semi-classical analysis. A discussion of the overall results and more thinks to look for, can be found in section $5$. Lastly, an appendix with some calculations is included.

\section{Classical description}

\subsection{The general solution}

The starting point is to provide the components of the general $3$-dim metric admitting a null Killing vector field. Let $\left(w,r,u\right)$ be the adopted to this field coordinates and an $2+1$ decomposition along the $r$ coordinate is put in use; then the components of the metric are given as follows \cite{Garcia-Diaz:2017cpv}
\begin{align}
g_{\grm\grn}=\begin{pmatrix}
0 & 0 & f(r,u)\\
0 & 1 & 0\\
f(r,u) & 0 & h(r,u)
\end{pmatrix}\label{1},
\end{align}
while the null Killing vector field is
\begin{align}
\grj^{\grm}=\left(1,0,0\right)\label{2}.
\end{align}
This paper is dedicated to the study of the more restricted family of geometries which can be reached if we demand that \eqref{2} is also covariantly constant; this new demand provides us with the further condition
\begin{align}
f(r,u)=\tilde{f}(u)\nonumber.
\end{align}
Under this assumption the metric \eqref{1} can be written as
\begin{align}
g_{\grm\grn}=\begin{pmatrix}
0 & 0 & 1\\
0 & 1 & 0\\
1 & 0 & h(r,u)
\end{pmatrix},\label{4}
\end{align}
where we have absorbed the function $\tilde{f}(u)$ with a coordinate transformation of the form $(w\rightarrow\tilde{w},r\rightarrow\tilde{r},u\rightarrow \tilde{u}=\int{\tilde{f}(u)du})$ and, for the sake of simplicity, we keep the same symbols. The metric \eqref{4} is called a pp-wave metric \cite{Kundt1961} and has the property that all it's first kind curvature scalars, such as $\left(R,R_{\grm\grn}R^{\grm\grn},R_{\grm\grn\grs\grr}R^{\grm\grn\grs\grr},...\right)$ are equal to zero \cite{Jordan2009}.  It's corresponding Einstein tensor $\left(G_{\grm\grn}\right)$ has only one non-vanishing component
\begin{align}
G_{33}=-\frac{1}{2}\partial_{rr}h(r,u),\nonumber
\end{align}
where with the symbol $\partial_{rr}$, a second partial derivative with respect to the variable $r$ is denoted and will be used wherever is needed from now on.

Let us now turn to the electromagnetic part of the system by giving the Faraday tensor in terms of the electric and magnetic fields,
\begin{align}
F_{\grm\grn}=\begin{pmatrix}
0 & -E_{1}(w,r,u) & -E_{2}(w,r,u)\\
E_{1}(w,r,u) & 0 & B_{1}(w,r,u)\\
E_{2}(w,r,u) & -B_{1}(w,r,u) & 0
\end{pmatrix}.\nonumber
\end{align}
The initial use of electric and magnetic fields instead of a three-potential $A_{\grm}\left(w,r,u\right)$ is made for the simplicity brought to the differential equations of the system. As we shall see, some conditions are easier to treat by using the fields. At this point we should clarify that the words and the symbols, electric (E) and magnetic (B) field, are used nominally, in the sense that we do not a priori know which of the coordinates will hold the role of ``time''. For more information on this subject take a look at \cite{2005CQGra..22..393T}.

It is well known, that the dynamics is described by the system of Einstein's-Maxwell's \footnote{The equations are presented in the Gaussian unit system and also we choose for simplicity $c=1$,\,$G=\frac{1}{2}$,\,$\hbar=1$, where $c$ the speed of light, $G$ the Newtonian gravitational constant and $\hbar$ the Planck's constant.} partial differential equations given in the following tensor-component form
\begin{align}
G_{\grm\grn}&=\,T_{\grm\grn},\label{7}\\
\nabla_{\grn}F^{\grm\grn}&=0,\label{8}\\
\nabla_{[\grs}F_{\grm\grn]}&=0,\label{9}
\end{align}
where the brackets $[]$ stand for total anti-symmetrization of the indices enclosed, and $T_{\grm\grn}$ is the corresponding electromagnetic energy-momentum tensor defined as
\begin{align}
T_{\grm\grn}=F_{\grm\grs}{F_{\grn}}^{\grs}-\frac{1}{4}g_{\grm\grn}F_{\grs\grr}F^{\grs\grr}\nonumber.
\end{align}
In order to simplify the form of the Faraday tensor and hence of the Einstein's-Maxwell's equations, some conditions may be applied. For the pp-wave spacetimes the following relation holds
\begin{align}
G_{\grm\grn}\,\grj^{\grm}\,\grj^{\grn}=0\xRightarrow{\eqref{7}}\,
T_{\grm\grn}\,\grj^{\grm}\,\grj^{\grn}=0\Rightarrow\,
E_{1}(w,r,u)=0.\nonumber
\end{align}
Next are the algebraic conditions which are relying on the vanishing components of the Einstein's tensor and the equation \eqref{7}, providing us with
\begin{align}
\forall i,j\,\, \text{with}\,\,i=j\neq{3},\,G_{ij}=0\xRightarrow{\eqref{7}}\,T_{ij}=0\Rightarrow\,
E_{2}(w,r,u)&=0.\nonumber
\end{align}
Finally, the Killing condition, recall it to be
\begin{align}
{\cal{L}}_{\xi}g_{\grm\grn}=0\Rightarrow\,{\cal{L}}_{\xi}G_{\grm\grn}=0\xRightarrow{\eqref{7}}\,
{\cal{L}}_{\xi}T_{\grm\grn}=0,\nonumber
\end{align}
where ${\cal{L}}_{\xi}$ denotes the Lie derivative, results in
\begin{align}
B_{1}(w,r,u)=B(r,u).\nonumber
\end{align}
In the light of these, the equations $\eqref{7},\eqref{8}$ become
\begin{align}
\frac{1}{2}\partial_{rr}h(r,u)+B(r,u)^{2}&=0,\nonumber\\
\partial_{r}B(r,u)&=0,\nonumber
\end{align}
while $\eqref{9}$ is identically satisfied. This system is easily integrated to give the general solution
\begin{align}
h(r,u)&=-r^{2}b(u)^{2}+r\,h_{2}(u)+h_{1}(u),\nonumber\\
B(r,u)&=b(u),\nonumber
\end{align}
where the functions $\left(b(u),h_{1}(u),h_{2}(u)\right)$ are completely arbitrary. It can be proven that, with the use of appropriate coordinate and electromagnetic gauge transformations, the functions $\left(h_{1}(u),h_{2}(u)\right)$ can be absorbed. It is thus safe to omit them altogether since there is no physical significance to which they are related. Note that for $b(u)=0$, $h_{1}(u)\neq{0},\,h_{2}(u)\neq{0}$ the system represents empty Minkowski spacetime. To conclude, the form of the general analytical solution is
\begin{align}
g_{\grm\grn}&=\begin{pmatrix}
0 & 0 & 1\\
0 & 1 & 0\\
1 & 0 & -r^{2}b(u)^{2}
\end{pmatrix},\label{18}\\
A_{\grm}&=(0,0,r\,b(u)),\label{19}
\end{align}
where $A_{\grm}$ is the corresponding electromagnetic potential and $b(u)$ some arbitrary function. We consider $F_{\grm\grn}=\nabla_{\grm}A_{\grn}-\nabla_{\grn}A_{\grm}$.

The question now arises about the nature of the arbitrary function $b(u)$; that is whether different such functions define different geometries.

For the purpose of answering the above question, it is more convenient to transform to new coordinate $\tilde{u}$, $d\tilde{u}=b(u)du$ and rename $\tilde{u}$ as $u$ for simplicity. The metric and the potential are then brought into the form
\begin{align}
g_{\grm\grn}&=\begin{pmatrix}
0 & 0 & f(u)\\
0 & 1 & 0\\
f(u) & 0 & -r^{2}
\end{pmatrix},\label{217}\\
A_{\grm}&=(0,0,r),\label{218}
\end{align}
That is to say, the function $b(u)$ was absorbed from the $g_{uu}$ component of \eqref{18} and is now represented by $f(u)=\frac{1}{b(u)}$. We have omitted a multiplicative absorb-able constant ($k_{0}$) in the $g_{uu}$ component since it's presence would only add the inclusion of the flat spacetime (for $k_{0}=0$). At the same time, the three-potential becomes $u$-independent and the corresponding Faraday tensor acquires one constant component, $F_{ru}=1$. In this form of the metric, we only have to check if another metric with a different $f_{1}(u)$ can be related by a coordinate transformation to \eqref{217}.

Let us start from \eqref{217} and assume a general transformation of the form
\begin{align}
w=h_{1}(w_{1},r_{1},u_{1}),\, r=h_{2}(w_{1},r_{1},u_{1}),\, u=h_{3}(w_{1},r_{1},u_{1}),\nonumber
\end{align}
and demand that the new metric be of the form
\begin{align}
g_{\grm\grn}(w_{1},r_{1},u_{1})&=\begin{pmatrix}
0 & 0 & f_{1}(u_{1})\\
0 & 1 & 0\\
f_{1}(u_{1}) & 0 & -r_{1}^{2}
\end{pmatrix},\nonumber
\end{align}
If the above transformation does indeed brings the two metrics in correspondence, it should also do the same for the Ricci tensors. But both of the Ricci tensors can be calculated to have only one non-vanishing component, namely $R_{uu}=1$. The two sets of equations
\begin{align}
g_{\grs\grt}(w_{1},r_{1},u_{1})=g_{\grm\grn}(w,r,u)\frac{\partial q^{\grm}}{\partial \tilde{q}^{\grs}}\frac{\partial q^{\grn}}{\partial \tilde{q}^{\grt}},\quad R_{\grs\grt}(w_{1},r_{1},u_{1})=R_{\grm\grn}(w,r,u)\frac{\partial q^{\grm}}{\partial \tilde{q}^{\grs}}\frac{\partial q^{\grn}}{\partial \tilde{q}^{\grt}},\nonumber
\end{align}
where $q^{\grm}=(w,r,u)$,\,$\tilde{q}^{\grs}=(w_{1},r_{1},u_{1})$, constrain the function $f_{1}$ to just a constant multiple of $f$; but the constant can be absorbed by a re-scaling of $w_{1}$. Therefore, the conclusion is that $f(u)$ is an essential function, that is every different $f(u)$ defines another geometry. This also means that a rather unusual situation occurs: a host of different geometries (parametrized by $f(u)$) are compatible with a constant Faraday tensor, i.e $F_{ru}=1$. The question as to the origin of this unusual finding is interesting; is it due to the absence of gravitational degrees of freedom in 3 dimensions or is it a property of the system (pp-wave + electromagnetic field)? Half of the answer lies in a different interpretation of the effective energy-momentum tensor associated to the given family of geometries \eqref{217}. Indeed, one can be satisfied that a mass-less scalar field can also serve as a source with $\phi(u)=u\equiv f^{-1}(f(u))$, and the only non-vanishing component of the energy momentum tensor to be $T_{uu}=(\partial_{u}\phi(u))^{2}$; thus $f(u)$ now appears both in the geometry and the matter content. The degeneracy is thus due to the electromagnetic field. The other half of the answer requires a relation to some $3+1$ model, where certainly there are gravitational degrees of freedom. Fortunately, there is the following embedding of the three dimensional classical system, above described, to a four dimensional one: as far as the metric is concerned, we envisage the trivial embedding, $ds_{(4)}^{2}=dt^{2}+2f(u)dwdu+dr^{2}-r^{2}du^{2}$, where $t$ is the third spatial coordinate. We also complement the trivial prolongations of the null covariantly constant vector field and the electromagnetic potential as $\grj_{(4)}=(0,1,0,0)$, $A^{(4)}_{\grm}=(0,0,0,r)$ respectively. It can be easily checked that $\grj_{(4)}$ continues to be null and covariantly constant Killing field indicating that the four dimensional geometry also represents a pp-wave. The 4D Faraday tensor still remains constant with the only non-zero component as above. Thus, although there are now gravitational degrees of freedom, the same degeneracy as in the 3D case occurs. In this case also, one can lift the degeneracy through the same interpretation of the effective energy momentum tensor as that of a mass-less scalar field. The conclusion is that the degeneracy is due to the form of the electromagnetic field and the presence of the pp-wave. Note that the very existence of an arbitrary function is a consequence of the nature of pp-waves, see the relevant discussion in the beginning of section $V$.

\subsection{Classification}

With the general solution at hand, we may proceed to a classification based on symmetry groups: that is the finding of the possible Killing vector fields of the metric \eqref{217} for the entire family designated by $f(u)$. In trying to do so some branches may appear, which may demand that the function $f(u)$ be of a specific form. As a result the cardinality, as well as the Type of the Lie algebra obeyed by the Killing vector fields may vary. In particular, two cases came up which are presented to some extent. For completeness, we recall the Killing-homothetic equation
\begin{align}
{\cal{L}}_{\grj}g_{\grm\grn}=\grv\,g_{\grm\grn},\nonumber
\end{align}
where $\grj$ is some general vector field and $\grv$ the homothetic parameter. The Classification will be based on the number of existing Killing vector fields. Nevertheless, we are also interested in the possible existence of a homothetic vector field: Firstly, because some non-essential constant may be absorbed through the motion induced by the corresponding integral curves. Secondly, since according to \cite{Edgar:2003rv}, if a homothetic and a gradient Killing vector field exist, then an irreducible Killing tensor field can be constructed, which may help for instance in the integration of the geodesic equations. \\
Starting from a general vector field $\grj$, we solve the Killing homothetic equations until we arrive at a point where their last components to be solved are
\begin{align}
c_{1}\left(\frac{\dot{f}(u)^{2}}{f(u)}-\ddot{f}(u)\right)=0,\label{khe1}\\
\dot{\grz}(u)-\frac{\dot{f}(u)}{f(u)}\grz(u)-\grz(u)^{2}-1=0,\label{khe2}
\end{align}
while the Killing-homothetic vector field is (up to this point)
\begin{align}
\grj=\left(c_{2}+w\,\grv+\frac{r}{f(u)}\grz(u)e^{c_{3}-\int{\grz(u)du}}-c_{1}w \frac{\dot{f}(u)}{f(u)},e^{c_{3}-\int{\grz(u)du}}+\frac{r\,\grv}{2},0\right).
\end{align}
The constants $c_{1},c_{2},c_{3},\grv$ are the parameters related to the Killing and homothetic vector field generators. The equation \eqref{khe2} being of first order will contribute one more parameter; there will thus be at most a four dimensional Killing-Lie algebra. When it comes to the equation \eqref{khe1}, there are two choices: either $c_{1}$ is zero, thus the Killing-Lie algebra will be of dimension three, or $c_{1}\neq{0}$ and the function $f(u)$ acquires a specific form.

\subsubsection{Case $I$, Four Killing vector fields}

As it is indicated by the title, this case arises when $c_{1}\neq{0}$ and hence \eqref{khe1} results
\begin{align}
f(u)=M e^{m u}.\nonumber
\end{align}
The constant $M$ can be absorbed from the metric \eqref{217} by a coordinate transformation of the form $w=\frac{\tilde{w}}{M}, r=\tilde{r}, u=\tilde{u}$. Based on the values of the constant $m$ the following Classes appear, which differ due to their Lie algebra:
\begin{center}
\begin{tabular}{ |c|c|c| }
\hline
\textbf{$f(u)$} & \textbf{Killing vector fields} & \textbf{Structure constants} \\
\hline
\multirow{2}{*}{$1$} & $\grj_{1}=(1,0,0)$,\,$\grj_{2}=(0,0,1)$, & $C^{4}_{23}=-C^{4}_{32}=-1,\,C^{3}_{24}=-C^{3}_{42}=1,$\\
& $\grj_{3}=(r sinu,cosu,0)$,\,$\grj_{4}=(-r cosu,sinu,0)$ & $C^{1}_{34}=-C^{1}_{43}=-1$\\
\hline
\multirow{3}{*}{$e^{2u}$} & $\grj_{1}=(1,0,0)$,\,$\grj_{2}=(-2w,0,1)$, & $C^{1}_{12}=-C^{1}_{21}=-2,\,C^{3}_{23}=-C^{3}_{32}=1,$\\
& $\grj_{3}=(-r(1+u)e^{-u},u e^{u},0)$, & $C^{4}_{23}=-C^{4}_{32}=1$,\,$C^{4}_{24}=-C^{4}_{42}=1,$\\
& $\grj_{4}=(-r e^{-u},e^{u},0)$ & $C^{1}_{34}=-C^{1}_{43}=1$\\
\hline
\multirow{3}{*}{$e^{mu},m\neq{0,2}$} & $\grj_{1}=(1,0,0)$,\,$\grj_{2}=(-mw,0,1)$, & $C^{1}_{12}=-C^{1}_{21}=-m$,\,$C^{3}_{23}=-C^{3}_{32}=\frac{m-\gra}{2}$,\\
& $\grj_{3}=(-\frac{r(m-\gra)}{2}e^{-1/2(m+\gra)u},e^{1/2(m-\gra)u},0)$, & $C^{4}_{24}=-C^{4}_{42}=\frac{(m+\gra)}{2}$,\,$C^{1}_{34}=-C^{1}_{43}=-\gra$\\
& $\grj_{4}=(-\frac{r(m+\gra)}{2}e^{-1/2(m-\gra)u},e^{1/2(m+\gra)u},0)$ & \\
\hline
\end{tabular}
\end{center}
where $\gra=\sqrt{-4+m^{2}}$. All the cases admit the same homothetic vector field $\grj_{h}=(w,\frac{r}{2},0)$. We have verified that there exist transformations of the Killing vector field basis, that map the obtained structure constants, to those of the following four dimensional Lie algebras $A_{(4,10)}$,\,$A_{(4,7)}$,\,$A_{(4,9)}^{b}$,\,with\,$0<\lvert b\rvert<1$, correspondingly. The constant $m$ should be redefined in terms of $b$ as $m=\frac{1+b}{\sqrt{b}}$. This enumeration of four dimensional Lie algebras can be found in J. Patera and P. Winternitz \cite{doi:10.1063/1.523441}, thus is safe to say that there exist three Classes.\\

\textbf{Remarks:}

\begin{enumerate}
\item All the metrics have Lorentzian signature, with the convention being $(+,+,-)$.
\item There is no restriction on the domain of the variables, $w,r,u\in \mathbb{R}$.
\item The variable $u$ holds the role of time, since the norm of the tangent vector to the lines $w=\text{constant}$,\,$r=\text{constant}$, e.g $v_{u}^{\grm}=(0,0,1)$, is timelike $\forall w,r,u\in \mathbb{R}$.
\item For all the three Classes, there exists only one simply transitive sub-group. This acts on the plane $(w,u)$ and consists of the first two Killing vector fields in each Class. The only difference being that in the first Class, the sub-algebra is Abelian, while non-Abelian in the other two.
\item To the best of our knowledge, these are new solutions except from the first Class. In a work of Clement \cite{0264-9381-10-5-002} a solution with the same physical content (e.g pp-wave electromagnetic spacetime) was found, by using a different method and prescribed in different coordinates $\left(t,\grr,\gru\right)$,
\begin{align}
g_{\grm\grn}&=\begin{pmatrix}
0 & 0 & \gra\\
0 & -\frac{1}{\gra^{2}} & 0\\
\gra & 0 & \frac{\gru_{0}}{m\,\gra^{2}}\left(\grr^{2}+c\right)
\end{pmatrix},\nonumber\\
A_{\grm}&=\left(0,0,\frac{\grp^{0}}{\gra}\grr\right),\nonumber
\end{align}
where $\left(\gra,\,\gru_{0},\,m,\,c,\,\grp^{0}\right)$ are constants. The constant m is not related with ours. A different convention was also used for the Lorentz signature $\left(+,-,-\right)$. Due to this, the coordinate transformation that connects those two metrics is imaginary as it is presented below,
\begin{align}
w=\gra\,\frac{\sqrt{m}}{\sqrt{\gru_{0}}}\,t+\frac{c}{2\,\gra^{2}}\frac{\sqrt{\gru_{0}}}{\sqrt{m}}\,\gru,\,
r=\frac{i}{\gra}\,\grr,\,
u=\frac{\sqrt{\gru_{0}}}{\sqrt{m}}\,\gru.\nonumber
\end{align}
Also, the following relation holds between the constants, $\gru_{0}=-m\,{(\grp^{0})}^{2}$,$\left(m<0\,\,\text{and}\,\,\grp^{0}>0 \right)$ in order to be aligned with our unit conventions.
\end{enumerate}

\subsubsection{Case $II$, Three Killing vector fields}

In this case, $c_{1}=0$ and the function $f(u)$ remains unspecified. The only restriction is that the previous forms corresponding to the three Classes (i.e. $f(u)\neq{1,e^{2u},e^{m u}}$) are excluded, since otherwise one of the Killing fields would have been missed. The last component of the Killing-homethetic equation to be solved is
\begin{align}
\dot{\grz}(u)-\frac{\dot{f}(u)}{f(u)}\grz(u)-\grz(u)^{2}-1=0.\label{220}
\end{align}
One can recognize \eqref{220} as a Ricatti equation. Up to this point, the general Killing-homothetic vector field has the form
\begin{align}
\grj=\left(c_{2}+w\,\grv+\frac{r}{f(u)}\grz(u)e^{c_{3}-\int{\grz(u)du}},e^{c_{3}-\int{\grz(u)du}}+\frac{r\,\grv}{2},0\right),\nonumber
\end{align}
where $c_{2},c_{3}$ are some parameters out of which two of the Killing generators will be produced and $\grv$ the homothetic parameter. Since \eqref{220} is a first order ordinary differential equation for $\grz(u)$, one constant of integration is expected. Thus, there will be a total of three integration constants corresponding to three symmetry generators, hence, indeed this is what designates this case from the previous one.

\subsection{Minisuperspace Lagrangian}

It is our intention to provide a minisuperspace Lagrangian description of the previously discussed system. In order to do so, we firstly recall that a necessary condition to be fulfilled for such a Lagrangian to exist, is that a
spacetime of dimension $(d+1)$ should admit a transitive(simply or multiply) symmetry Lie group, acting on the hyper-surfaces of dimension $d$. In the previous section we started from a metric admitting only one covariantly constant field hence, in particular, a Killing field. Of course, on ``mass shell'', more Killing fields appeared and enumerated in the classification. Only the Classes of case $I$ were found to admit a simply transitive two dimensional group, which acts on the $r=\text{constant}$ hyper-surface. Thus, we may expect to succeed in finding an appropriate minisuperspace description for these Classes. We could equivalently have started from a metric admitting these simply transitive groups and reproduce the solutions exhibited in the previous section. For each Class the group is different. We observe that in the Class $A_{(4,10)}$ there are the Killing fields, $\left\{\grj_{1}=(1,0,0),\grj_{2}=(0,0,1)\right\}$ which form an Abelian sub-algebra and hence an Abelian sub-group, while in the Classes $A_{(4,7)}$,\,$\left\{\grj_{1}=(1,0,0),\grj_{2}=(-2w,0,1)\right\}$ and $A_{(4,9)}^{b}$,\,$\left\{\grj_{1}=(1,0,0),\grj_{2}=(-m w,0,1)\right\}$ the sub-algebra is non-Abelian. Due to the \textit{simple transitive} action of the previous groups, these correspond to the two existing two-dimensional ``Bianchi Types''. In the sections to follow, we present the form of the metrics and the three-potentials. The reasoning for acquiring such a form can be found in Appendix $A$.

\subsection{Class $A_{(4,10)}$}

The metric components of the three-dimensional spacetime in coordinates $\left(z,t,\grt\right)$ and the three-potential, acquire the form
\begin{align}
\tilde{g}_{\grm\,\grn}&=\begin{pmatrix}
0 & 0 & \tilde{\grv} \\
0 & n(t)^{2} & 0 \\
\tilde{\grv} & 0 & c(t)\\
\end{pmatrix},\label{69}\\
\tilde{A}_{\grm}&=\left(0,0,A(t)\right),\label{70}
\end{align}
where $\tilde{\grv}$ is some constant.
Let us write down the system of Einstein's-Maxwell's equations and discuss about an important aspect concerning them.
\begin{align}
\ddot{c}(t)&=-2\,\dot{A}(t)^{2}+\dot{c}(t)\,\frac{\dot{n}(t)}{n(t)},\label{71}\\
\ddot{A}(t)&=\dot{A}(t)\,\frac{\dot{n}(t)}{n(t)}\label{72}.
\end{align}
where the dot ($\cdot$) represents derivative with respect to $t$. There are only two second order ordinary differential equations and no constraint, which one might expect to exist due to the remaining freedom of arbitrarily re-parameterizing the variable $t$. In other words $n(t)$ is not a dynamical degree of freedom(i.e. it is a lapse function); yet there is no corresponding constraint equation. It is seems like this is a characteristic of the pp-wave nature of the geometry. By choosing $n(t)=1$ we find the solution
\begin{align}
A(t)&=\grm_{1}+\grm_{2}\,t,\label{73}\\
c(t)&=-\grm_{2}^{2}\,t^{2}+\grm_{3}\,t+\grm_{4},\label{74}
\end{align}
where $\grm_{1},\grm_{2},\grm_{3},\grm_{4}$ are integration constants; if we perform the following transformation, the metric \eqref{69} is cast into the form found in the previous section,
\begin{align}
z=\frac{\grm_{2}}{\tilde{\grv}}w-\frac{{\grm_{3}}^{2}+4\,\grm_{4}\,\grm_{2}^{2}}{8\,\grm_{2}^{3}\,\tilde{\grv}}u,\,
t=r+\frac{\grm_{3}}{2\,\grm_{2}^{2}},\,
\grt=\frac{1}{\grm_{2}}u.\nonumber
\end{align}
Furthermore, the potential \eqref{218} is reached with the aid of the above transformation and electromagnetic gauge transformation acting upon \eqref{70}
\begin{align}
A_{\grm}=\tilde{A}_{\grn}\frac{\partial x^{\grn}}{\partial y^{\grm}}+\nabla_{\grm}\grL,\nonumber
\end{align}
where $\grL$ is
\begin{align}
\grL(w,r,u)=-\frac{2\grm_{1}\grm_{2}+\grm_{3}}{2\grm_{2}^{2}}\,u,\nonumber
\end{align}
and $x^{\grn}=(\grz,t,\grt)$,\,$y^{\grm}=(w,r,u)$. We verify that the solution can be reproduced in this way.

Let us now turn to our original desire to find a minisuperspace Lagrangian. The usual way of seeking such a reduced Lagrangian is to insert the metric \eqref{69} and the potential \eqref{70} into the Einstein's-Hilbert's+Electromagnetic action. This is not working here due to a fact that, as we have already discussed, in the class of pp-wave spacetimes all the curvature scalars are zero.

Nevertheless, this problem can be circumvented by trying to construct a Lagrangian by inspection of the equations \eqref{71},\eqref{72}. At first we observe that there is no ``potential'' part hence the Lagrangian will consist of only a ``kinetic'' term. There are two dynamical degrees of freedom $\left(c(t),A(t)\right)$ related to the two second order ordinary differential equations which are linear in the second derivative. Thus, the ``kinetic part'' will have to be quadratic in the ``velocities'' $\left(\dot{c}(t),\dot{A}(t)\right)$. Additionally, since there is only a first derivative of $n(t)$ in the equations, the Lagrangian should depend on at most
$n(t)$. These facts allow us to write the generic form
\begin{align}
{\cal{L}}=\tilde{G}_{i\,j}\left(n,\,c,\,A\right)\dot{q}^{i}\,\dot{q}^{j},\nonumber
\end{align}
where $q^{i}=(c,\,A)$, while the $t$-dependence has been suppressed for simplicity. At this point, we may also take advantage of our previous knowledge concerning the Lagrangians for constraint systems, and arrive at the final form
\begin{align}
{\cal{L}}=\frac{1}{2\,n}G_{i\,j}\left(c,A\right)\dot{q}^{i}\,\dot{q}^{j}.\nonumber
\end{align}
With this Lagrangian at hand, we are able to reproduce \eqref{71},\eqref{72} via the Euler-Lagrange equations if and only if the minisuperspace metric $G_{i\,j}$ has the form
\begin{align}
G_{i\,j}=\begin{pmatrix}
0 & 1 \\
1 & 4\,A \\
\end{pmatrix}.\label{82}
\end{align}
A constraint equation will also appear, which is related to the variation with respect to $n$, and reads
\begin{align}
\dot{c}+2\,A\,\dot{A}=0.\label{83}
\end{align}
When the solutions found above are inserted into \eqref{83} the following relation between constants results
\begin{align}
\grm_{2}\,\left(2\,\grm_{1}\,\grm_{2}+\grm_{3}\right)=0.\label{84}
\end{align}

This looks like an apparent discrepancy. The supplementary constraint equation \eqref{83} seems to impose extra restriction on the integration constants involved in the solutions \eqref{73},\eqref{74}. A possible way out would be, to treat $n$ as a non-dynamical degree of freedom but rather as an t-dependent $1/ ``\text{mass}''$ term and therefore not perform a variation of the Lagrangian with respect to it. Once treated like this, the resulting Euler-Lagrange equations are exactly \eqref{71},\eqref{72}.

On the other hand, even if $n(t)$ is considered dynamically and thus arrive at \eqref{83}, condition \eqref{84} will not have any effect upon the physical content of the solutions: since there are no essential constants in the geometry of this Class, all the constants appearing are absorbable via coordinate and electromagnetic gauge transformations and thus the constraint relation will always be satisfied. It is seems like the two ways of treating $n$ result eventually to the same solution. Perhaps, the reason behind this is the non-existence of essential constants. Let us see if something similar will come up in the next Classes as well.

\subsubsection{Classes $A_{(4,7)}$,\,$A_{(4,9)}^{b}$}

Due to the non-Abelian nature of the group in these two Classes, we may treat them both at once. Thus, we introduce the parameter $\tilde{m}$ with respect to which the two Classes are parametrized. For different specific values of this parameter, the two Classes are retrieved. We recall that the metric admitting this non-Abelian group, is separated into the two Classes $A_{(4,7)}$\,$A_{(4,9)}^{b}$ only after the solutions have been found. The steps followed are the same as previously. Thus, in order to avoid to be tedious, we present only the important features. We provide the components of the 3-D metric and the electromagnetic potential.\\

\begin{align}
\tilde{g}_{\grm\grn}&=\begin{pmatrix}
0 & 0 & e^{\tilde{m}\grt}\tilde{\grv}\\
0 & n(t)^{2} & 0 \\
e^{\tilde{m}\grt}\tilde{\grv} & 0 & c(t)
\end{pmatrix},\,\tilde{m}\neq{0},\nonumber\\
\tilde{A}_{\grm}&=(0,0,A(t)).\nonumber
\end{align}

As it turns out, the Einstein's-Maxwell's equations for both Classes, are the same with the ones of the Class $A_{(4,10)}$ given by \eqref{71},\eqref{72}. Hence, the same solutions for $c(t),A(t)$ will come up. The coordinate and electromagnetic gauge transformations that transform the solutions into the form given in section II.2.1 are,
\begin{align}
z=\frac{\grm_{2}}{\tilde{\grv}}w+\frac{\grm_{3}^{2}+4\grm_{4}\grm_{2}^{2}}{8\grm_{2}^{2}\tilde{\grv}\tilde{m}}e^{-\frac{\tilde{m}}{\grm_{2}}u},\,t&=r+\frac{\grm_{3}}{2\grm_{2}^{2}},\,\grt=\frac{u}{\grm_{2}},\nonumber\\
\grL(w,r,u)&=-\frac{2\grm_{1}\grm_{2}+\grm_{3}}{2\grm_{2}^{2}}\,u,\nonumber
\end{align}
where we have used again the freedom to set $n(t)=1$.
The Class $A_{(4,7)}$ is acquired when $\tilde{m}=2\grm_{2}$, while the Class $A_{(4,9)}^{b}$ when $\tilde{m}=m\grm_{2}$,\,$m\neq{0,2}$.

Since the equations are the same, the Lagrangian capable to incorporate the dynamics is the same as in the previous Class. Once more, the constraint equation does not affect the physical content of the solutions. None of the constants appearing in the constraint was essential, even though $A_{(4,9)}^{b}$ has one essential constant, $m$.

Another fact is also worth mentioning: if we compare the electromagnetic gauge scalars $\grL$ of all the Classes, and the ``on mass shell'' constraint \eqref{84}, it is easy to see that there is a common factor, namely
\begin{align}
2\,\grm_{1}\,\grm_{2}+\grm_{3}.\nonumber
\end{align}
When a solution of this equation is provided or, in other words, the constraint is solved, it seems like there is no need for electromagnetic gauge transformation, the coordinate transformation suffices to bring $\tilde{A}_{\grm}$ into \eqref{218}. Perhaps, this is another possible explanation of why the solutions are physically equivalent independently of the constraint; the constraint is incorporated completely in the electromagnetic gauge transformation.

Finally, as we saw, there is an equivalence between the two ways of treating $n(t)$ at the classical level. It would be interesting to see whether this equivalence holds also in the quantum regime.

\subsection{Minisuperspace Hamiltonian}

Let us now construct the minisuperspace Hamiltonian for the three Classes of the first case. As we have said, there is only one Lagrangian for all the Classes and hence there will be only one Hamiltonian. However, the construction of the Hamiltonian depends on whether we treat $n(t)$ as dynamical or non-dynamical degree of freedom. Both cases will be treated. As a preliminary step, let us point out some facts about the minisuperspace metric which will provide us with a simpler description of the system.

It is easy to verify that \eqref{82} is flat and has Lorentz signature. Subsequently, there is a coordinate transformation which transforms the metric into the standard Minkowski form:
\begin{align}
c=-\left(y+x\right)^{2}-\frac{1}{2}\left(y-x\right),\,
A=y+x.\nonumber
\end{align}
The Lagrangian is now
\begin{align}
{\cal{L}}=\frac{1}{2\,n}\left(-\dot{y}^{2}+\dot{x}^{2}\right),\nonumber
\end{align}
where $(\cdot)$ stands for derivative with respect to t. Since the space is flat, the following Killing vector fields exist
\begin{align}
\grz_{1}=\left(1,0\right),\,
\grz_{2}=\left(0,1\right),\,
\grz_{3}=\left(x,y\right),\nonumber
\end{align}
where the first argument signifies the $y$ direction, satisfying the Lie algebra
\begin{align}
\left[\grz_{1},\grz_{2}\right]_{L}=0,\,
\left[\grz_{2},\grz_{3}\right]_{L}=\grz_{1},\,
\left[\grz_{3},\grz_{1}\right]_{L}=-\grz_{2}.\nonumber
\end{align}
The corresponding Euler-Lagrange equations are
\begin{align}
\dot{y}^{2}-\dot{x}^{2}=0,\,
\ddot{y}=\frac{\dot{n}}{n}\,\dot{y},\,
\ddot{x}=\frac{\dot{n}}{n}\,\dot{x}.\nonumber
\end{align}
where $n$ was treated as a dynamical variable. If not, the first one would not exist.

In the next sections, the Hamiltonian description will be presented.

\subsection{Non-dynamical $n(t)$}

Since $n(t)$ is treated as a non-dynamical degree of freedom, the Lagrangian is \textit{regular} and we proceed in finding the canonical Hamiltonian as usual, by calculating the canonical momenta of the dynamical variables:
\begin{align}
p_{y}=-\frac{\dot{y}}{n},\,
p_{x}=\frac{\dot{x}}{n}.\nonumber
\end{align}
By solving these equations with respect to the velocities we obtain
\begin{align}
\dot{y}=-n\,p_{y},\,
\dot{x}=n\,p_{x},\nonumber
\end{align}
while the canonical Hamiltonian is calculated via the Legendre transformation
\begin{align}
{\cal{H}}_{can}=\frac{n}{2}\left(-p_{y}^{2}+p_{x}^{2}\right).\nonumber
\end{align}
The time evolution of some quantity $B$ in the phase space are given as
\begin{align}
\frac{d\,B}{d t}=\frac{\partial\,B}{\partial t}+\left\{B,{\cal{H}}_{can}\right\},\nonumber
\end{align}
where $B=B(t,y,x,p_{y},p_{x})$. Thus, the phase space equations of motion are
\begin{align}
\dot{y}=-n\,p_{y},\,
\dot{x}=n\,p_{x},\,
\dot{p}_{y}=0,\,
\dot{p}_{x}=0.\nonumber
\end{align}
Due to the explicit $t$ dependence, the canonical Hamiltonian is not conserved in general,
\begin{align}
\frac{d{\cal{H}}_{can}}{dt}=\frac{1}{2}\left(-p_{y}^{2}+p_{x}^{2}\right)\dot{n}\equiv{\cal{H}}_{can}\frac{\dot{n}}{n}.\nonumber
\end{align}
On the other hand, due to Noether's theorem, each of the Killing vector fields is related to a conserved charge, which in our case reads
\begin{align}
Q_{1}=p_{y},\,
Q_{2}=p_{x},\,
Q_{3}=x\,p_{y}+y\,p_{x}.\nonumber
\end{align}
These conserved charges will prove useful in the Quantum description.

\subsection{Dynamical $n(t)$}

For $n(t)$ considered as a dynamical degree of freedom, the Lagrangian is \textit{singular}. This been taken into account, the known procedure for this kind of systems must be employed.

The canonical momenta are defined as always
\begin{align}
p_{y}=-\frac{\dot{y}}{n},\,
p_{x}=\frac{\dot{x}}{n},\,
p_{n}=0.\nonumber
\end{align}
The canonical Hamiltonian is now defined via the Legendre transformation
\begin{align}
{\cal{H}}_{can}=\frac{n}{2}\left(-p_{y}^{2}+p_{x}^{2}\right),\label{125}
\end{align}
while a primary constraint exists
\begin{align}
\grF_{1}\coloneqq p_{n}.\nonumber
\end{align}
The \eqref{125} is not capable of reproducing the equations of motion for this system, thus the total Hamiltonian needs to be defined
\begin{align}
{\cal{H}}_{tot}\coloneqq{\cal{H}}_{can}+\grl\,\grF_{1},\nonumber
\end{align}
where $\grl$ is some Lagrange multiplier which remains unspecified on the classical orbits; the constraint must satisfy the ``weak equality''
\begin{align}
\grF_{1}\approx{0},\nonumber
\end{align}
weak meaning that we must first calculate all the Poisson brackets and then apply the constraint. There is a further consistency condition for the state of system to remain in the constraint surface:
\begin{align}
\dot{\grF}_{1}\coloneqq\left[\grF_{1},{\cal{H}}_{tot}\right]_{P}\approx{0},\nonumber
\end{align}
which results in a secondary constraint
\begin{align}
\grF_{2}\coloneqq\frac{1}{2}\left(-p_{y}^{2}+p_{x}^{2}\right).\nonumber
\end{align}
The conservation of this constraint is identically satisfied; thus no further constraint exist. These two are named \textit{first class} according to Dirac, due to the following property
\begin{align}
\left[\grF_{1},\grF_{2}\right]_{P}\approx{0}.\nonumber
\end{align}
In the light of these facts, the total Hamiltonian may be expressed as a linear combination of the first class constraints
\begin{align}
{\cal{H}}_{tot}=n\,\grF_{2}+\grl\,\grF_{1}.\nonumber
\end{align}
This Hamiltonian correctly reproduces the equations of the system.
The two constraints are conserved as well as the charges mentioned in the previous section.

\section{Quantum description}

For the purposes of the present work, the canonical quantization procedure will be adopted. The canonical momenta and positions are promoted to operators $\left(p_{j}\rightarrow \hat{p}_{j}, x^{j}\rightarrow \hat{x}^{j}\right)$ which satisfy the following properties
\begin{enumerate}
\item \textbf{self-adjoint}
\begin{align}
\left<\hat{p}_{j}\grc\lvert\grf\right>=\left<\grc\lvert\hat{p}_{j}^{*}\grf\right>,\,
\left<\hat{x}^{j}\grc\lvert\grf\right>=\left<\grc\lvert(\hat{x}^{j})^{*}\grf\right>,\nonumber
\end{align}
with $\left<\grc\lvert\grf\right>=\int{d^{2}x\,\grm\,\grc^{*}\,\grf}$ the inner product and $\grm=\sqrt{\lvert det\left(G_{jl}\right)\rvert}$ the measure \cite{1986NCimB..93....1C}.
\item \textbf{canonical commutation relations}
\begin{align}
\left[\hat{x}^{j},\hat{x}^{l}\right]_{C}=0,\,
\left[\hat{x}^{j},\hat{p}_{l}\right]_{C}=i\,\grd^{j}_{l},\,
\left[\hat{p}_{j},\hat{p}_{l}\right]_{C}=0,\nonumber
\end{align}
\end{enumerate}
where $\left[\cdot,\cdot\right]_{C}$ denotes the commutator and $\grd^{j}_{l}$ the Kronecker's delta. The position representation suffices for the previous properties to hold
\begin{align}
\hat{p}_{j}=-i\,\partial_{j},\,
\hat{x}^{j}=x^{j}.\nonumber
\end{align}

\subsection{Non-dynamical $n(t)$}

As it is well known, the transition from the classical to the quantum regime requires the resolution of the factor ordering problem. This resolution is better to based on the properties of the classical system. The scaling of the lapse $n\rightarrow \tilde{n}\,\grv^{-1}(y,x)$ is a covariance of the classical system when n is considered dynamical and thus the Hamiltonian is weakly zero.  Therefore the Hamiltonian operator should also be covariant under the above scaling which points to the choice of the conformal Laplacian or Yamabe operator,
\begin{align}
\hat{\cal{H}}_{can}&=-n\left(\frac{1}{2}\frac{1}{\grm}\partial_{j}\left(\grm G^{jl}\partial_{l}\right)-\frac{d-2}{8(d-1)}\mathcal{R}\right).\nonumber
\end{align}
For the case when $n$ is not considered dynamical and thus the Hamiltonian is not weakly zero, the above argumentation is not applicable. However, for the sake of uniformity we can also choose this operator as well. Note that, in the present case, the final form of the operators is just that of the Laplace-Beltrami since both the space is flat, $\mathcal{R}=0$, and two dimensional.
As far as the conserved charges are concerned, the corresponding linear operators are
\begin{align}
\hat{Q}_{I}&=-i\,\left[\grz_{I}^{j}\partial_{j}+\frac{1}{2\,\grm}\partial_{j}\left(\grm\,\grz^{j}_{I}\right)\right],\nonumber
\end{align}
where $(I=1,2,3)$ distinguishes indices internal to the algebra. Note also that on account of $\grz_{I}$ being Killing fields, the second term of the linear quantum operators also vanishes.
It is easy to verify that
\begin{align}
\left[\hat{\cal{H}}_{can},\hat{Q}_{I}\right]_{C}\grc=0,\,
\left[\hat{Q}_{1},\hat{Q}_{2}\right]_{C}\grc=0,\,
\left[\hat{Q}_{2},\hat{Q}_{3}\right]_{C}\grc=-i\,\hat{Q}_{1}\grc,\,
\left[\hat{Q}_{3},\hat{Q}_{1}\right]_{C}\grc=i\,\hat{Q}_{2}\grc,\nonumber
\end{align}
thus there are two distinct maximal operator sub-algebras of interest. The first is $\left(\hat{Q}_{1},\hat{Q}_{2},\hat{\cal{H}}_{can}\right)$ and the other $\left(\hat{Q}_{3},\hat{\cal{H}}_{can}\right)$.

The most interesting fact in this case, related to the non-dynamical nature of $n(t)$ and hence to the Lagrangian being \textit{regular}, is the possibility to define a Schrodinger-like instead of the usual Wheeler-DeWitt equation for singular systems. This is not common at all, since in the most cases (e.g. cosmology or point sources), if not all, the minisuperspace Lagrangian is \textit{singular} and some specific reductions are employed in order to end up with a Schrodinger-like equation. As a result of this interesting aspect of this system, the quantum states to be found will be subtend to evolution with respect to $t$. We assume that the variable $t$, which here is to be understood as a parameter, is in one to one correspondence with the parameter-time appearing in the quantum description of non-relativistic point particles. Thus, the system of equations to be solved for each subalgebra is
\begin{align}
i\,\partial_{t}\grc&=\hat{\cal{H}}_{can}\grc,\label{146}\\
\hat{Q}_{I}\grc&=p_{I}\grc,\label{147}
\end{align}
where $p_{\grn}$ the eigenvalues. Note that \eqref{146}, unlike the usual Schrodinger's equation, is invariant under reparametrizations $t\rightarrow\tilde{t}=\tilde{t}(t)$; this is due to the transformation law of $n(t)$ (inherited by it's position in the line element) and the fact that the partial derivative on the right hand side transforms also in the same way. As a consequence, the results should not depend on the choice of $n(t)$.
In the following table, the solutions of \eqref{146},\eqref{147} are presented.

\begin{center}
\begin{tabular}{ |c|c| }
\hline
\textbf{Sub-algebra} & \textbf{Eigenstate}\\
\hline
$\left(\hat{Q}_{1},\hat{Q}_{2},\hat{\cal{H}}_{can}\right)$ & $\grc_{12}\left(t,y,x\right)=\frac{1}{2\pi}e^{-i E_{12}\int{n(t)\,dt}}\,e^{i\left(p_{1}y+p_{2}x\right)}$\\
\hline
$\left(\hat{Q}_{3},\hat{\cal{H}}_{can}\right)$ & $\grc_{3}\left(t,\grs,\gru\right)=e^{-i E_{3}\int{n(t)dt}}e^{i p_{3}\gru}\left[q_{1}J_{i p_{3}}(-i \sqrt{2 E_{3}}\grs)+q_{2}Y_{i p_{3}}(-i \sqrt{2 E_{3}}\grs)\right]$\\
\hline
\end{tabular}
\end{center}

where $E_{12}=\frac{1}{2}\left(-p_{1}^{2}+p_{2}^{2}\right)$, and the eigenstates $\grc_{12}$ where normalized by use of the Dirac's delta function
\begin{align}
\left<\grc_{p_{1}p_{2}}(y,x)\lvert\grc_{\tilde{p}_{1}\tilde{p}_{2}}(y,x)\right>=\grd\left(p_{1}-\tilde{p}_{1}\right)\grd\left(p_{2}-\tilde{p}_{2}\right).\nonumber
\end{align}
Also, $J_{\grn}(z),Y_{\grn}(z)$ represent the Bessel functions of the first and second kind respectively, $q_{1},q_{2}$ are integration constants and $E_{3}$ the eigenvalue of the $t$-independent Schrodinger's equation. The coordinates $\grs,\gru$, are those which transform the operator $\hat{Q}_{3}$ into it's normal form.
\begin{align}
y=\grs Cosh\gru,\,x=\grs Sinh\gru,\Leftrightarrow\grs=\sqrt{y^{2}-x^{2}},\,\gru=ArcTanh\left(\frac{x}{y}\right),\,
G_{jl}=\begin{pmatrix}
-1 & 0\\
0 & \grs^{2}\\
\end{pmatrix}.\nonumber
\end{align}
It is important to note that the solutions $\grc_{12},\grc_{3}$ hold because $\hat{\cal{H}}_{can}$ commutes with itself at different $t$, else one would need to introduce time ordering in the exponential.\\
Some comments regarding the orthogonality of $\grc_{3}$ can be found in Appendix $B$.

\subsection{Dynamical $n(t)$}

For the dynamical case we follow the Dirac-Bergmann algorithm, meaning that the constraint operators should result zero when acting upon physical states, $\hat{\grF}_{J}\grc=0$. The set of operators consists of
\begin{align}
\hat{\grF}_{2}&=-\frac{1}{2}\frac{1}{\grm}\partial_{j}\left(\grm G^{jl}\partial_{l}\right)+\frac{d-2}{8(d-1)}\mathcal{R},\nonumber\\
\hat{\grF}_{1}&=-i \partial_{n},\nonumber\\
\hat{Q}_{I}&=-i\,\left[\grz_{I}^{j}\partial_{j}+\frac{1}{2\,\grm}\partial_{j}\left(\grm\,\grz^{j}_{I}\right)\right].\nonumber
\end{align}
The solutions are presented below,

\begin{center}
\begin{tabular}{ |c|c| }
\hline
\textbf{Sub-algebra} & \textbf{Eigenstate}\\
\hline
$\left(\hat{Q}_{1},\hat{Q}_{2},\hat{\grF}_{1},\hat{\grF}_{2}\right)$ & $\grc_{12}\left(n,y,x\right)=\frac{1}{2 \pi}e^{i\left(p_{1}y+p_{2}x\right)}$\\
\hline
$\left(\hat{Q}_{3},\hat{\grF}_{1},\hat{\grF}_{2}\right)$ & $\grc_{3}\left(n,\grs,\gru\right)=e^{i p_{3}\gru}\left(q_{1}e^{i p_{3}\grs}+q_{2}e^{-i p_{3}\grs}\right)$\\
\hline
\end{tabular}
\end{center}

where the equation
\begin{align}
-p_{1}^{2}+p_{2}^{2}=0\label{222}
\end{align}
should hold and $\grc_{12}$ was normalized as previously. Once more, for the same reasons as before, in the second sub-algebra the following coordinate transformation was performed,
\begin{align}
y=e^{\grs}Cosh\gru,\,x=e^{\grs}Sinh\gru,\Leftrightarrow\grs=\frac{1}{2}Log(y^{2}-x^{2}),\,\gru=ArcTanh\left(\frac{x}{y}\right),\,
G_{jl}=e^{2\grs}\begin{pmatrix}
-1 & 0\\
0 & 1\\
\end{pmatrix}.\nonumber
\end{align}

For more details on the normalization of $\grc_{3}$ take a look at Appendix $B$.

\section{Bohm analysis}

We have found the analytical expression of the wavefuctions for all the cases. Due to our difficulty in interpreting these wave functions in terms of the usual quantum mechanics, we use the Bohm analysis \cite{PhysRev.85.166,PhysRev.85.180,Bohm1984} which provides deterministic trajectories (i.e. geometries). This analysis incorporates the notion of the classical trajectory and provides us with the tools to compare the quantum corrected (semiclassical) metrics with the classical ones. In order to get a glimpse of those quantum corrections to the classical solutions, we choose to compare the solutions of $\left(\hat{Q}_{1},\hat{Q}_{2},\hat{\cal{H}}_{can}\right)$ and $\left(\hat{Q}_{1},\hat{Q}_{2},\hat{\grF}_{1},\hat{\grF}_{2}\right)$. Let us first recall in a few lines the mathematical background needed: Suppose that a system is described by the following Hamiltonian operator
\begin{align}
\hat{\cal{H}}=-\gra \nabla_{j}\nabla^{j}+V,\nonumber
\end{align}
and eigenstates which can be cast into the form
\begin{align}
\grc=\grV e^{i S},\label{176}
\end{align}
where $V(x^{j})$ is some potential, $\grV(x^{j})$ is called the amplitude and $S(x^{j})$ is the phase of the eigenstate. By use of \eqref{176} in Schrodinger's equation, the imaginary part provide us with a sort of continuity equation
\begin{align}
\frac{\partial \grV}{\partial \grr}+2\gra \nabla^{j}\grV\nabla_{j}S+\gra \grV\nabla_{j}\nabla^{j}S=0,\nonumber
\end{align}
while the real part with a Hamilton-Jacobi-like equation
\begin{align}
\frac{\partial S}{\partial \grr}+\gra \nabla_{j}S\nabla^{j}S+V-Q=0,\nonumber
\end{align}
where $Q=\gra \frac{\nabla_{j}\nabla^{j}\grV}{\grV}$, and is called the quantum potential, since is the term which comes as a new entry in the usual Hamilton-Jacobi equation. Following Bohm, the connection to the classical level can be achieved by assuming that the definition of $\nabla_{j}S$ is to be the momentum of the system. Thus, the identification with canonical momenta leads to the following system of equations
\begin{align}
\nabla_{j}S=\frac{\partial {\cal{L}}}{\partial \dot{x}^{j}}.\label{179}
\end{align}
In the absence of quantum potential should be expected for the classical solutions to be retrieved.

\subsection{Non-dynamical $n(t)$}

\subsubsection{Eigenstate}

The eigenstate $\grc_{12}$ is already in the desired form, thus the amplitude and the potential can be read off
\begin{align}
\grV=\frac{1}{2\pi},\,
S=p_{1}y+p_{2}x-\frac{1}{2}(-p_{1}^{2}+p_{2}^{2})\int{n(t)\,dt}.\nonumber
\end{align}
Since, $\grV$ is constant, the quantum potential is zero.

The solutions to \eqref{179} are easily found to be
\begin{align}
y(t)=\grn_{1}-p_{1}\int{n(t)dt},\,
x(t)=\grn_{2}+p_{2}\int{n(t)dt},\label{250}
\end{align}
with $\grn_{1},\grn_{2}$ integration constants. Transform back to the degrees of freedom $c,f$, and by use of $n(t)=1$, the three Classes are retrieved by mere of the following coordinate and electromagnetic gauge transformations
\begin{enumerate}
\item Class $A_{(4,10)}$
\begin{align}
z&=\frac{p_{2}-p_{1}}{\tilde{\grv}}w+\frac{p_{2}^{2}\left(1-16\grn_{1}\right)+p_{1}^{2}\left(1+16\grn_{2}\right)+2p_{1}p_{2}\left(1+8\grn_{1}-8\grn_{2}\right)}{32\tilde{\grv}\left(p_{1}-p_{2}\right)^{3}}u,\nonumber\\
t&=r+\frac{(p_{2}+p_{1})-4(p_{2}-p_{1})(\grn_{1}+\grn_{2})}{4(p_{1}-p_{2})^{2}},\,
\grt=\frac{1}{p_{2}-p_{1}}u,\nonumber\\
\grL&=-\frac{p_{1}+p_{2}}{4\left(p_{1}-p_{2}\right)^{2}}u.\nonumber
\end{align}
\item Classes $A_{(4,7)}$,\,$A_{(4,9)}^{b}$
\begin{align}
z&=\frac{p_{2}-p_{1}}{\tilde{\grv}}w+\frac{p_{2}^{2}\left(1-16\grn_{1}\right)+p_{1}^{2}\left(1+16\grn_{2}\right)+2p_{1}p_{2}\left(1+8\grn_{1}-8\grn_{2}\right)}{32\tilde{\grv}\tilde{m}\left(p_{1}-p_{2}\right)^{2}}e^{-\tilde{m}u/(p_{2}-p_{1})},\nonumber\\
t&=r+\frac{(p_{2}+p_{1})-4(p_{2}-p_{1})(\grn_{1}+\grn_{2})}{4(p_{1}-p_{2})^{2}},\,\grt=\frac{1}{p_{2}-p_{1}}u,\nonumber\\
\grL&=-\frac{p_{1}+p_{2}}{4\left(p_{1}-p_{2}\right)^{2}}u,\nonumber
\end{align}
with $\tilde{m}=2\left(p_{2}-p_{1}\right)$ for $A_{(4,7)}$ and $\tilde{m}=m\left(p_{2}-p_{1}\right)$ for $A_{(4,9)}^{b}$,\,$m\neq{0,2}$.
\end{enumerate}
The choice $p_{1}=p_{2}$ in \eqref{250} would result the flat space.

As expected, due to the absence of quantum potential the semi-classical trajectories coincide with the classical. A question arises; is it possible to acquire a different solution from the classical ones? The answer is \textbf{yes}, but we have to deviate form the eigenstate.\\

\subsubsection{Gaussian initial state}

We assume that the initial state of the system is a Gaussian distribution (arising from a superposition of pure states)
\begin{align}
\grF_{0}(y,x)=\sqrt{\frac{\grl}{\grp}}e^{-\frac{\grl}{2}\left(y^{2}+x^{2}\right)},\nonumber
\end{align}
with it's evolution given by
\begin{align}
\grc(t,y,x)=\int_{-\infty}^{\infty}{\int_{-\infty}^{\infty}{c(p_{1},p_{2})\grF_{p_{1}p_{2}}}(y,x)e^{-i E_{12} \int{n(t)dt}}}dp_{1}dp_{2},\label{193}
\end{align}
where
\begin{align}
\grF_{p_{1}p_{2}}(y,x)=\frac{1}{2\grp}e^{i p_{1} y}e^{i p_{2} x},\,
c(p_{1}p_{2})=\int_{-\infty}^{\infty}{\int_{-\infty}^{\infty}{\grF^{*}_{p_{1}p_{2}}\grF_{0}(y,x)dydx}}.\nonumber
\end{align}
The parameter $\grl$ determines the concentration of the initial state around the point $(y,x)=(0,0)$. The larger the value the higher the concentration. After a lot of algebraic manipulations the desired form of the state \eqref{193} is found, with the amplitude and phase being
\begin{align}
\grV=\sqrt{\frac{\grl}{\grp}}\frac{1}{\sqrt{1+\grl^{2}B(t)^{2}}}exp\left[{-\frac{\grl}{2\left(1+\grl^{2}B(t)^{2}\right)}\left(y^{2}+x^{2}\right)}\right],\,
S=\frac{\grl^{2}B(t)}{2\left(1+\grl^{2}B(t)^{2}\right)}\left(x^{2}-y^{2}\right),\nonumber
\end{align}
where $B(t)=\int{n(t)dt}$. The quantum potential is now not zero
\begin{align}
Q=\frac{n(t)\left(-y^{2}+x^{2}\right)\grl^{2}}{2\left[1+\grl^{2}\left(\int{n(t)dt}\right)^{2}\right]^{2}}.\nonumber
\end{align}
The same steps as before are performed and the solutions for $n(t)=1$ are
\begin{align}
y(t)=\grn_{1}\sqrt{1+t^{2}\grl^{2}},\,
x(t)=\grn_{2}\sqrt{1+t^{2}\grl^{2}}.\nonumber
\end{align}
As a result, the following semi-classical fields came up
\begin{enumerate}
\item Abelian(Bianchi Type I)
\begin{align}
g_{\grm\grn}=\begin{pmatrix}
0 & 0 & 1\\
0 & 1 & 0\\
1 & 0 & -r^{2}\left(1+\grs\frac{\sqrt{1+r^{2}\grl^{2}}}{r^{2}\grl^{2}}\right)\\
\end{pmatrix},\,
A_{\grm}=\left(0,0,\frac{\sqrt{1+r^{2}\grl^{2}}}{\grl}\right),\label{202}
\end{align}
where $\grs=\frac{\grn_{1}-\grn_{2}}{2\left(\grn_{1}+\grn_{2}\right)^{2}}$, and the following coordinate transformations were performed
\begin{align}
z=\frac{\grl(\grn_{1}+\grn_{2})}{\tilde{\grv}}w+\frac{\grn_{1}+\grn_{2}}{2\grl\tilde{\grv}}u,\,t=r,\,\grt=\frac{1}{\grl(\grn_{1}+\grn_{2})}u.\nonumber
\end{align}

The branch $\grn_{1}+\grn_{2}=0$, leads to
\begin{align}
g_{\grm\grn}=\begin{pmatrix}
0 & 0 & 1\\
0 & 1 & 0\\
1 & 0 & \sqrt{1+r^{2}\grl^{2}}\\
\end{pmatrix},\,
A_{\grm}=\left(0,0,0\right),\label{225}
\end{align}
after the transformation
\begin{align}
z=\frac{\sqrt{\grn_{2}}}{\tilde{\grv}}w,\,t=r,\,\grt=\frac{1}{\sqrt{\grn_{2}}}u.\nonumber
\end{align}
\item Non-Abelian(Bianchi Type II)
\begin{align}
g_{\grm\grn}=\begin{pmatrix}
0 & 0 & e^{\tilde{m}u/\grl(\grn_{1}+\grn_{2})}\\
0 & 1 & 0\\
e^{\tilde{m}u/\grl(\grn_{1}+\grn_{2})} & 0 & -r^{2}\left(1+\grs\frac{\sqrt{1+r^{2}\grl^{2}}}{r^{2}\grl^{2}}\right)\\
\end{pmatrix},\,
A_{\grm}=\left(0,0,\frac{\sqrt{1+r^{2}\grl^{2}}}{\grl}\right),\label{204}
\end{align}
where the corresponding transformations are
\begin{align}
z=\frac{\grl(\grn_{1}+\grn_{2})}{\tilde{\grv}}w-\frac{(\grn_{1}+\grn_{2})^{2}}{2\tilde{m}\tilde{\grv}}e^{-\tilde{m}u/\grl(\grn_{1}+\grn_{2})},\,t=r,\,\grt=\frac{1}{\grl(\grn_{1}+\grn_{2})}u.\nonumber
\end{align}

Once again, the branch $\grn_{1}+\grn_{2}=0$ leads to
\begin{align}
g_{\grm\grn}=\begin{pmatrix}
0 & 0 & e^{\tilde{m}u/\sqrt{\grn_{2}}}\\
0 & 1 & 0\\
e^{\tilde{m}u/\sqrt{\grn_{2}}} & 0 & \sqrt{1+r^{2}\grl^{2}}\\
\end{pmatrix},\,
A_{\grm}=\left(0,0,0\right),\label{240}
\end{align}
with
\begin{align}
z=\frac{\sqrt{\grn_{2}}}{\tilde{\grv}}w,\,t=r,\,\grt=\frac{1}{\sqrt{\grn_{2}}}u.\nonumber
\end{align}
\end{enumerate}

\textbf{Remarks:}

\begin{enumerate}
\item The metric \eqref{202} admits only the original two Killing fields of the Abelian Bianchi Type, $\grj_{1}=(1,0,0)$,\,$\grj_{2}=(0,0,1)$, and no homothecy. The same holds for \eqref{204} but with the non-Abelian Bianchi Type, $\grj_{1}=(1,0,0)$,\,$\grj_{2}=\left(-\frac{\tilde{m}w}{\grl(\grn_{1}+\grn_{2})},0,1\right)$.
\item Both of the solutions \eqref{202},\eqref{204}, do not satisfy the source-less Einstein's-Maxwell's equations. The existence of a three-current and an additional energy-momentum tensor of the following form has to be considered for both cases,
\begin{align}
T_{\grm\grn}=\begin{pmatrix}
0 & 0 & 0\\
0 & 0 & 0\\
0 & 0 & \frac{2\sqrt{1+r^{2}\grl^{2}}+\grs}{2\left(1+r^{2}\grl^{2}\right)^{3/2}}\end{pmatrix},\,
J_{\grm}=\left(0,0,-\frac{\grl}{4\grp(1+r^{2}\grl^{2})^{3/2}}\right).\label{206}
\end{align}
\item The two branches that appear, \eqref{225},\eqref{240}, are characterized by the absence of electromagnetic field, yet, they are not the flat spacetime as we might expect. In fact, they satisfy Einstein's equations with an energy momentum tensor of the form
\begin{align}
T_{\grm\grn}=\begin{pmatrix}
0 & 0 & 0\\
0 & 0 & 0\\
0 & 0 & \frac{\grl^{2}}{2(1+r^{2}\grl^{2})^{3/2}}\end{pmatrix}.\nonumber
\end{align}
\end{enumerate}

\subsection{Dynamical $n(t)$}

\subsubsection{The sub-algebra $\hat{Q}_{1},\hat{Q}_{2},\hat{\grF}_{1},\hat{\grF}_{2}$}

The eigenstate is already in the required form, with the amplitude and phase being
\begin{align}
\grV=\frac{1}{2\pi},\,
S=p_{1}y+p_{2}x.\nonumber
\end{align}
It turns out that the solutions are exactly the same as those described by  \eqref{250}. If we now take into account the condition \eqref{222}, we have the flat spacetime for $p_{1}=p_{2}$. For $p_{1}=-p_{2}$, the classical solutions in the previously given form, are retrieved via the following coordinate and electromagnetic gauge transformations
\begin{enumerate}
\item Class $A_{(4,10)}$
\begin{align}
z=\frac{2p_{2}}{\tilde{\grv}}w+\frac{\grn_{1}-\grn_{2}}{8p_{2}}u,\,
\grr=r-\frac{\grn_{1}+\grn_{2}}{2p_{2}},\,
\grt=\frac{1}{2p_{2}}u,\,
\grL=0.\nonumber
\end{align}
\item Classes $A_{(4,7)}$,\,$A_{(4,9)}^{b}$
\begin{align}
z=\frac{2p_{2}}{\tilde{\grv}}w-\frac{\grn_{1}-\grn_{2}}{4\tilde{m}\tilde{\grv}}e^{-\tilde{m}u/2p_{2}},\,
\grr=r-\frac{\grn_{1}+\grn_{2}}{2p_{2}},\,
\grt=\frac{1}{2p_{2}}u,\,
\grL=0.\nonumber
\end{align}
with $\tilde{m}=4p_{2}$ for $A_{(4,7)}$ and $\tilde{m}=2m p_{2}$,$m\neq{0,2}$, for $A_{(4,9)}^{b}$.
\end{enumerate}
In the absence of quantum potential, the semi-classical and classical trajectories coincide. Unlike the previous case, the use of another initial state like e.g. another Gaussian will not result in a form required by the \eqref{176}, in order for the Bohm analysis to begin. This is due to the absence of ``evolution'' in constrained systems (frozen in time picture). Thus no quantum potential can be attained within this method. This answers to the negative the question of equivalence between the two ways of treating $n(t)$ at the quantum level.

\section{Discussion}

In this paper we have initially investigated the classical three-dimensional electromagnetic pp-wave spacetimes. We have obtained the entire solution space. To this end, we first implement the $2+1$ decomposition of the spacetime along a spatial coordinate and the Gauss normal coordinates ($N^{i}=0,N=1$) was also put in use. The second step was to apply the conditions which are necessary for a spacetime to represent a pp-wave. Next, a set of conditions, emanating from the symmetries of the problem, has been applied to the electromagnetic energy momentum tensor; as a result the set of Einstein's-Maxwell's equations were reduced to a set of two partial differential equations. Their solution was easily found and the interesting fact is the existence of an arbitrary essential function appearing in the final form of the fields. Thus, in a coordinate system in which the electromagnetic field is constant, there are infinitely many compatible geometries. To the best of our knowledge, this is at variance to what happens in usual cosmological or point-like situations where particular electromagnetic configurations determine the corresponding geometries up to constants (see e.g. the Reisner-Nordstrom solution). In order to explain this a bit further, let us consider the following: for the case of linearized Einstein's equations (in which $h_{ij}$ below is considered $|h_{ij}|<<1$ ), the line element of plane waves propagate along the $z$ direction, can be cast into the form
\begin{align}
ds^{2}=-dt^{2}+dz^{2}+\left(\grd_{ij}+h_{ij}(z-t)\right)dy^{i}dy^{j},\label{fm}
\end{align}
where $h_{ij}(z-t)$ are arbitrary functions, $(t,z,y^{i})$ the coordinates of the spacetime, $\grd_{ij}$ the metric of the flat Euclidean space and the index $i$ may acquire any countable value, thus \eqref{fm} is meant for any dimension. In the case of vacuum solutions, the matrix $h_{ij}$ should be traceless, while in the non-vacuum case, some other kind of condition will be imposed associated to the matter source chosen. The arbitrariness of this function is related to the linearity of the equations and the fact that wave solutions can be superimposed. Let us use the light cone coordinates $U=\frac{1}{\sqrt{2}}(z-t)$, $V=\frac{1}{\sqrt{2}}(z+t)$ which transform the line element into the form
\begin{align}
ds^{2}=2dUdV+\left(\grd_{ij}+h_{ij}(\sqrt{2}U)\right)dy^{i}dy^{j}.\nonumber
\end{align}
If the assumption about $h_{ij}$ being small is dropped and we could find a solution of the form
\begin{align}
ds^{2}=2dUdV+\tilde{g}_{ij}(U)dy^{i}dy^{j},\label{fm1}
\end{align}
then we would say that \eqref{fm1} represents a plane wave solution of the non linear Einstein's equations. The coordinates $(U,V,y^{i})$ are the so called Rosen coordinates. If by any chance, the functions $\tilde{g}_{ij}(U)$ remained arbitrary, the superimposition of the waves would still exist in the non-linear regime. All that we have said up to now, is valid for any spacetime dimension.

When it comes to our case, the general solution \eqref{18} was presented in what is called the Brinkmann coordinates
\begin{align}
ds^{2}=2dwdu-r^{2}b(u)^{2}du^{2}+dr^{2}.\nonumber
\end{align}
With a coordinate transformation of the form $w=V-\frac{y^{2}M(U)}{2}Exp[2\int{M(U)dU}]$, $r=y Exp[\int{M(U)dU}]$, $u=U$ the line element transforms into the Rosen coordinates
\begin{align}
ds^{2}=2dUdV+Exp[2\int{M(U)dU}]dy^{2},\label{fm2}
\end{align}
if and only if, the function $M(U)$ satisfies the following Ricatti equation
\begin{align}
\dot{M}(U)+M(U)^{2}+b(U)^{2}=0.\label{Rce}
\end{align}
The \eqref{fm2} has the same form with \eqref{fm1} for $i=1,j=1$. The arbitrariness of $b(U)$ is carried over to the metric \eqref{fm2} through the solutions of the equation \eqref{Rce}. Thus, the physical significance of the arbitrariness of the function $b(u)$ is reflected to the capability of superimposing strong gravito-electromagnetic pp-waves. Due to the absence of true gravitational degrees of freedom, if $b(u)$ were zero (thus no electromagnetic field), then the solution would correspond to flat spacetime. As we can see, the arbitrariness of the function is a characteristic of the wave nature of the solutions and is expected to exist in any dimension. In contrast, the superposition of point-like solutions is not again a solution. Let us provide particular examples of some $b(U)$ for which the superposition of the solutions holds. In the following table we present the specific forms of the functions $b(U), M(U)$ in three cases. Furthermore we provide the term $Exp[2\int{M(U)dU}]$ of the line element \eqref{fm2}. The chosen functions $b(U), M(U)$ satisfy the equation \eqref{Rce} and provide a specific example of superposition of waves.

\begin{center}
\begin{tabular}{ |c|c|c| }
\hline
$b(u)$ & $M(u)$ & $Exp[2\int{M(U)dU}]$ \\
\hline
$\sqrt{2 sec(2U)^{2}-tan(2U)^{2}}$ & $-tan(2U)$ & cos(2U)\\
\hline
$\sqrt{2csc(2U)^{2}-cot(2U)^{2}}$ & $ cot(2U)$ & sin(2U)\\
\hline
$\sqrt{\frac{3+sin(4U)}{1+sin(4U)}}$ & $\frac{cos(2U)-sin(2U)}{cos(2U)+sin(2U)}$ & $cos(2U)+sin(2U)$\\
\hline
\end{tabular}
\end{center}
Note however that the above superposition does not hold for the electromagnetic potential, e.g. in the Rosen coordinates reads
\begin{align}
A_{\grm}=(0,0,y\,b(U)\,Exp[\int{M(U)dU}]),\nonumber
\end{align}
where the previous statement is quite obvious.
To conclude this paragraph, we expect arbitrary functions to exist when we consider pp-wave spacetimes, in any dimension.

The idea to classify the solution space based on the symmetry group proved to be fruitful. Specifically, two cases occurred: \\
Case $I$ was divided in three Classes where the previously arbitrary function acquired a specific form in each Class separately. The Killing fields in each Class were explicitly determined and their cardinality is four. \\
Case $II$, the essential function remained arbitrary while the number of Killing fields is reduced to three. Their explicit form is not given, but we managed to pin point the unknown part into the solution of a single Riccati equation. For each explicit form of the essential function, the Riccati should be solved and then the explicit form of the Killing fields will be obtained.

For the three Classes of case $I$, we are able to reproduce the solutions based on a minisuperspace Lagrangian. This Lagrangian is based on a three dimensional configuration space. The variation of the corresponding action with respect to the two ``dynamical'' degrees of freedom leads to the only two second order ordinary differential equations describing the system. The third degree of freedom is the lapse $n(t)$; when it is considered as dynamical ( leading to a singular Lagrangian), a constraint equation occurs, hence a difference between the number of the  Euler-Lagrange and the system's equations. When $n(t)$ is considered as non-dynamical (point to a regular Lagrangian), but rather as a $t$-dependent term, no such discrepancy appears. Nevertheless, the discrepancy is only nominal since the solutions in both cases are equivalent. Thus, it seems that the solutions are unaffected by the way we treat $n(t)$. In trying to explain this, we presented two possible reasons: The first is based on the absence of essential constants in the constraint, when it was calculated ``on mass shell''. The second has to do with the complete absorption of the constraint in the electromagnetic gauge transformation.

In order to shed some light on the physical significance of the absence of constraint equations regarding the metrics $II.4$, $II.4.1$, it is instructive to perform some coordinate transformations. Recall that the initial coordinates are $(z,t,\grt)$ where $t$ is a spatial coordinate, while the null vector field providing us with the direction of propagation of the wave is $\grj=\partial_{z}$. Consider the following transformations in the $(II.4)$ and non-abelian case $(II.4.1)$ respectively.
\begin{align}
&z=-\frac{1}{2\tilde{\grv}}\left[\left(u-y\right)+\left(u+y\right)c(x)\right],\,\,t=x,\,\,\grt=u+y,\,\,\, \text{$(II.4)$}\nonumber\\
&z=\frac{1}{2}\left[-\left(u-y\right)+\frac{c(x)}{\tilde{m}^{2}\left(u+y\right)}\right],\,\,t=x,\,\,\grt=-\frac{1}{\tilde{m}}Log\left[\frac{\tilde{\grv}}{\tilde{m}\left(u+y\right)}\right],\,\,\, \text{$(II.4.1)$}\nonumber
\end{align}
where the new coordinates are $(u,x,y)$ with $u$ the time-leke coordinate. The coordinate $x$ is in one to one correspondence with the coordinate $t$, thus the form of Einstein's-Maxwell's equations are not altered, neither the form of the Lagrangian capable of incorporating the dynamics. The line element and the electromagnetic potential of $(II.4)$ acquire the form
\begin{equation}
\begin{aligned}\nonumber
&ds_{(2+1)}=n(x)^{2}dx^{2}-(u+y)\dot{c}(x)\,dx\,du-(u+y)\dot{c}(x)\,dx\,dy-du^{2}+dy^{2},\\
&A=A(x)\,du+A(x)\,dy,
\end{aligned}
\end{equation}
while for $(II.4.1)$
\begin{equation}
\begin{aligned}\nonumber
&ds_{(2+1)}=n(x)^{2}dx^{2}+\frac{\dot{c}(x)}{\tilde{m}^{2}\left(u+y\right)}\,dx\,du+\frac{\dot{c}(x)}{\tilde{m}^{2}\left(u+y\right)}\,dx\,dy-du^{2}+dy^{2},\\
&A=\frac{A(x)}{\tilde{m}\left(u+y\right)}\,du+\frac{A(x)}{\tilde{m}\left(u+y\right)}\,dy.
\end{aligned}
\end{equation}
In the spirit of the $(2+1)$ decomposition of the metric along the spatial coordinate $x$, we infer the existence of a lapse function $n(x)$ in both cases, the shift vector fields $N=-\frac{\left(u+y\right)\dot{c}(x)}{2}du-\frac{\left(u+y\right)\dot{c}(x)}{2}dy$ and $N=\frac{\dot{c}(x)}{2\tilde{m}^{2}\left(u+y\right)}du+\frac{\dot{c}(x)}{2\tilde{m}^{2}\left(u+y\right)}dy$, and the line element of the $x=constant$ surfaces, $ds_{(2)}=-du^{2}+dy^{2}$. The shift vector fields are null with respect to the internal metric of the surface and furthermore, they are the dynamical variables of the system in these coordinates. Additionally, the wave propagation vector field becomes $\grj=-\partial_{u}+\partial_{y}$. Let us recall the constraint equations in $(2+1)$ analysis:
\begin{align}
&G_{00}=\frac{1}{2}\left(-R^{(2)}+K^{2}-K_{ij}K^{ij}\right),\nonumber\\
&G_{0i}=D_{i}K-D_{j}{K_{i}}^{j},\nonumber
\end{align}
where $R^{(2)}$, $K_{ij}$, $K$ the Ricci scalar of the $x=constant$ surfaces, the extrinsic curvature tensor and it's trace correspondingly, while $D_{i}$ the covariant derivative compatible with the internal metric. The index $i$ refers to the coordinates $(u,y)$ of the surfaces. The tensor $K_{ij}$ is expressed in terms of the metric and the shift vector field
\begin{align}
K_{ij}=\frac{1}{2n(x)}\left(\partial_{x}h_{ij}-D_{i}N_{j}-D_{j}N_{i}\right),\nonumber
\end{align}
with the metric being
\begin{align}
h_{ij}=\begin{pmatrix}
-1 & 0 \\
0 & 1 \end{pmatrix},\nonumber
\end{align}
From the form of the metric we can easily infer the internal flatness of the surfaces, $R^{(2)}=0$, in both cases. Thus, the surfaces $x=constant$ are two dimensional flat Minkowski manifolds. The extrinsic curvature tensors have the following form in the two cases,
\begin{align}
K_{ij}=\frac{\dot{c}(x_{0})}{2n(x_{0})}\begin{pmatrix}
1 & 1 \\
1 & 1 \end{pmatrix},\,\,
K_{ij}=\frac{\dot{c}(x_{0})}{2\tilde{m}^{2}\left(u+y\right)^{2}n(x_{0})}\begin{pmatrix}
1 & 1 \\
1 & 1 \end{pmatrix},\nonumber
\end{align}
calculated at some surface $x=x_{0}$ after performing the derivative.

The information about the embedding of those flat surfaces is provided by the extrinsic curvature and we have easily calculate that $K=0$, $K_{ij}K^{ij}=0$, ${D_{j}K_{i}}^{j}=0$. Under transformations of $(u,y)$ coordinates the first two quantities transform as scalars while the third as a vector field, thus will be zero in the transformed system. The equation $K=0$ states that a congruence of curves with tangent vector field the unit normal to the surfaces, will be neither converging nor diverging. Thus, the extrinsic curvature tensor acts like a traceless shear tensor on curves upon the surfaces. The equation $K_{ij}K^{ij}=0$ can hold only in the case of Lorentzian manifolds and since the surface has dimension $2$, the components of $K_{ij}$ are equal as can be observed from the above expressions. A more visualized approach on the effect of shear parameters can be found in \cite{poisson2004relativist}. Finally, the equation $D_{j}{K_{i}}^{j}=0$ can be seen as the decomposition of the extrinsic curvature tensor in two diverge-less vector fields, ${K_{1}}^{j}$, ${K_{2}}^{j}$. A congruence of light curves, in the surfaces $x=constant$, generated by the propagation wave vector, are not affected by the shear.

To conclude, the absence of constraint equations states that the 3D manifold can be understand as the foliation of two dimensional flat timelike surfaces, along the spacelike coordinate $x$. Those surfaces are parallel to each other due to $K=0$. As a physical significance we deduce that an observer standing at some spatial distance $x=x_{0}$ will observe the wave propagating in the $y$ direction with the speed of light. Therefore, the regular Lagrangian is able to reproduce the dynamics of the system, describing the ``evolution'' of the two dimensional flat Minkowski surfaces along the $x$ direction.

The existence of the arbitrary lapse function even if there is no constraint equation is related to the general covariance of the theory. As it known, if Einstein's+matter equations where solved in their generality, with the use of appropriate boundary conditions, $D$(the dimension of the spacetime) arbitrary functions would appear in the metric. These $D$ functions correspond to the freedom of choosing a coordinate system. In our case, the symmetries of the problem and the geometry of the surfaces, are such that the constraint equations were satisfied identically without the need to specify all the arbitrary functions a priori. Since, it seems to be a property of the pp-wave spacetimes it might appear in higher dimensions as well.

In the quantum regime, we employed the canonical quantization for both the regular and singular Lagrangian. In order to interpret the solutions for one of the operator sub-algebras, we have used the Bohm analysis. As we have found, only when the Lagrangian was regular and the state of the system was a superposition of the eigenstates (specifically we study the evolution of a Gaussian initial state) a non-zero quantum potential exists and thus the semi-classical trajectories are deviating from the classical. As a result, the equivalence appearing at the classical level between the two ways of treating $n(t)$, is ``broken'' at the quantum level.

Furthermore, the non-trivial semi-classical solutions found do not satisfy the source-less Einstein's-Maxwell's equations; a three-current and an additional energy-momentum tensor must be considered. Both of them vanish for large values of the coordinate $r$,\,$r\rightarrow \infty$, which corresponds to a widely spread wavepacket compared to the original state. Note also that at this limit, the solution \eqref{202} coincides with Class $A_{(4,10)}$, while \eqref{204} coincides with Class $A_{(4,7)}$ when $\tilde{m}=2\grl(\grn_{1}+\grn_{2})$ and with $A_{(4,9)}^{b}$ when $\tilde{m}=m\grl(\grn_{1}+\grn_{2})$,\,$m\neq{0,2}$. Thus, the important result is that, even though we have considered one of the simplest possible initial quantum states, all three Classes are contained in the semi-classical solutions.

Another interesting fact that we have observed is the following: When $n(t)$ was treated non-dynamically, the Hamiltonian of the system had the same form like the one used to describe the motion of a free particle with variable mass in two dimensions. But there is an important difference; the Schrodinger's equation was invariant under the transformation $t\rightarrow f(\tilde{t})$ and this is due to the transformation law of $n(t)$. This is not the case for the variable mass system, since there the mass transforms as a scalar. With this caveat in mind, since we have found the Hilbert space in the case of the Gaussian initial state, we are able to write a probability density,
\begin{align}
P(t,y,x)=\frac{\grl}{\grp\left[1+\grl^{2}\left(\int{n(t)dt}\right)^{2}\right]}\exp\left[-\frac{\grl\left(y^{2}+x^{2}\right)}{1+\grl^{2}\left(\int{n(t)dt}\right)^{2}}\right],\nonumber
\end{align}
and to verify that indeed, even in the quantum description, the results does not depend on the choices of $n(t)$.

Lastly, we would like to point out some ideas for future work.
\begin{enumerate}
\item Construction of more complicated initial states for the algebra $\left(\hat{Q}_{1},\hat{Q}_{2},\hat{\cal{H}}_{can}\right)$, and their semi-classical analysis.
\item Construction of a midisuperspace Lagrangian capable to incorporate the dynamics of Case $II$ as well. Possible quantum description.
\item Analysis of the corresponding four dimensional system.
\end{enumerate}

\begin{appendices}
\section{2D ``Bianchi'' Types}

In the context of canonical formalism \cite{PhysRev.116.1322} the line element of a ($2+1$)-dimensional manifold $M$, in coordinates $(\gry,y^{i})$, $i=1,2$, acquires the form
\begin{align}
{ds^{2}_{(2+1)}}=\left(n^{2}+N_{i}N^{i}\right)\,d\gry^{2}+2\,N_{i}\,dy^{i}\,d\gry+\grg_{ij}\,dy^{i}\,dy^{j},\label{555}
\end{align}
where $n(\gry,y^{l}), N_{i}(\gry,y^{l})$ are the lapse function and shift vector components respectively and $\grg_{ij}(\gry,y^{l})$ the metric components of the $2$-dimensional sub-manifold which is given by $\gry=constant$.

Since we are interested in a manifold $M$ which admits a $2$-dimensional isometry group $G$, which acts simply transitively on the $2$-dimensional sub-manifold $\gry=constant$, we know that there exists an invariant basis of one-forms $\{\sigma^{\alpha}\}$ satisfying the curl relations \cite{1975hrc..book.....R}
\begin{align}
d\sigma^{\alpha}=-\frac{1}{2}C^{\alpha}_{\beta\,\epsilon}\,\sigma^{\beta}\wedge\sigma^{\epsilon}\Leftrightarrow \partial_{i}\sigma^{\alpha}_{j}-\partial_{j}\sigma^{\alpha}_{i}=-C^{\alpha}_{\beta\,\epsilon}\,\sigma^{\beta}_{i}\,\sigma^{\epsilon}_{j},\nonumber
\end{align}
where the Greek indices run from 1 to 2 and $C^{\alpha}_{\beta\epsilon}$ are the structure constants of the Lie algebra of the isometry group. The sub-manifold is then called homogeneous.
Under this assumption, a coordinate system with coordinates $(t,x^{i})$ exists such that the line element \eqref{555} acquires the manifestly homogeneous form
\begin{align}
{ds^{2}_{(2+1)}}=\left[n(t)^{2}+N_{\alpha}(t)N^{\alpha}(t)\right]\,dt^{2}+2\,N_{\alpha}(t)\sigma^{\alpha}_{i}(x^{l})\,dx^{i}\,dt+\gamma_{\alpha\beta}(t)\,\sigma^{\alpha}_{i}(x^{l})\,\sigma^{\beta}_{j}(x^{l})\,dx^{i}\,dx^{j}.\nonumber
\end{align}
It was proven in \cite{2001JMP....42.3580C} that coordinate transformations that preserve the sub-manifold's manifest homogeneity exist, such that the shift vector can always be set equal to zero in the transformed system, hence the final form of the line element is given by
\begin{align}
{ds^{2}_{(2+1)}}=n(t)^{2}dt^{2}+\gamma_{\alpha\beta}(t)\sigma^{\alpha}_{i}(x^{l})\sigma^{\beta}_{j}(x^{l})dx^{i}dx^{j}.\label{a2}
\end{align}
Let us now proceed with the different classes of case $I$.

\subsection{``Bianchi'' Type $I$}

The coordinates $x^{i}$ of \eqref{a2} are chosen to be labeled as $\left(z,\grt\right)$. The components of the one-forms for the Abelian case are
\begin{align}
\grs^{\gra}_{i}&=\begin{pmatrix}
1 & 0 \\
0 & 1 \\
\end{pmatrix}.\nonumber
\end{align}
If the matrix components $\grg_{\gra\grb}$ are given by
\begin{align}
\grg_{\gra\,\grb}&=\begin{pmatrix}
a(t) & b(t) \\
b(t) & c(t) \\
\end{pmatrix},\nonumber
\end{align}
then the metric components of the three-dimensional space in coordinates $\left(z,t,\grt\right)$ acquires the form
\begin{align}
\tilde{g}_{\grm\,\grn}=\begin{pmatrix}
a(t) & 0 & b(t) \\
0 & n(t)^{2} & 0 \\
b(t) & 0 & c(t)\\
\end{pmatrix}.\nonumber
\end{align}
This form looks the same with the one which someone usually starts from, in order to reproduce the BTZ black hole \cite{PhysRevD.90.024052}.
As we have said in the introduction of section II.3, we assumed that one of the Killing vector fields, specifically $\grj_{1}=(1,0,0)$, is null and covariantly constant, which leads to
\begin{align}
\tilde{g}_{\grm\,\grn}=\begin{pmatrix}
0 & 0 & \tilde{\grv} \\
0 & n(t)^{2} & 0 \\
\tilde{\grv} & 0 & c(t)\\
\end{pmatrix},\nonumber
\end{align}
where $\tilde{\grv}$ is some constant, while the electromagnetic potential which respects the symmetries is
\begin{align}
\tilde{A}_{\grm}=\left(0,0,A(t)\right).\nonumber
\end{align}

\subsection{``Bianchi Type'' $II$}
Once more, the Killing field which we demand to be null and covariantly constant is $\grj_{1}=(1,0,0)$.
The components of the one-forms for the non-Abelian case are
\begin{align}
\grs^{\gra}_{i}&=\begin{pmatrix}
e^{\tilde{m}\grt} & 0 \\
0 & 1 \\
\end{pmatrix},\,\tilde{m}\neq{0}.\nonumber
\end{align}
The same steps as before are followed, resulting
\begin{align}
\tilde{g}_{\grm\,\grn}=\begin{pmatrix}
0 & 0 & e^{\tilde{m}\grt}\tilde{\grv}  \\
0 & n(t)^{2} & 0 \\
e^{\tilde{m}\grt}\tilde{\grv}  & 0 & c(t)\\
\end{pmatrix},\tilde{A}_{\grm}=(0,0,A(t)).\nonumber
\end{align}

\section{Orthogonality of the wave functions}

\subsection{Non-dynamical $n(t)$, eigenstate $\grc_{3}$}

Recall the eigenstate
\begin{align}
\grc_{3}\left(t,\grs,\gru\right)=e^{-i E_{3}\int{n(t)dt}}e^{i p_{3}\gru}\left[q_{1}J_{i p_{3}}(-i \sqrt{2 E_{3}}\grs)+q_{2}Y_{i p_{3}}(-i \sqrt{2 E_{3}}\grs)\right],\nonumber
\end{align}
where $\grs\in(0,+\infty)$,\,$\gru\in(-1,1)$.
The differential equation
\begin{equation}
  \grs \frac{d}{d \grs}\left(\grs \frac{d F}{d\grs}\right) - (a^2 \grs^2 - k^2) F =0,\nonumber
\end{equation}
has the general solution
\begin{equation}
  F = C_1 J_{i k} (i a \grs) + C_2 Y_{i k} (i a \grs),\nonumber
\end{equation}
with $a\in \mathbb{R}$ being the eigenvalue of the equation. Let us choose a solution belonging to the class spanned by $F$, which we denote with $f_{k,a}$. Of course, it satisfies the same equation
\begin{equation} \label{eqa}
  \grs \frac{d}{d \grs}\left(\grs \frac{d f_{k,a}}{d\grs}\right) - (a^2 \grs^2 - k^2) f_{k,a} =0.
\end{equation}
Given that $\grs, k \in \mathbb{R}$, we take the complex conjugate of the above equation, but consider a different eigenvalue than before which we denote with $b \in \mathbb{R}$ and we have
\begin{equation} \label{eqb}
  \grs \frac{d}{d \grs}\left(\grs \frac{d \barr{f_{k,b}}}{d\grs}\right) - (b^2 \grs^2 - k^2) \barr{f_{k,b}} =0.
\end{equation}
Multiplication of \eqref{eqa} with $\barr{f_{k,b}}$, \eqref{eqb} with $f_{k,a}$ and substraction of the two resulting relations leads to
\begin{equation}
  \frac{d}{d \grs} \left[ \grs \left(\barr{f_{k,b}} \frac{d}{d\grs}f_{k,a} - f_{k,a} \frac{d}{d\grs}\barr{f_{k,b}} \right)\right] - (a^2-b^2) \grs \barr{f_{k,b}} f_{k,a} = 0.\nonumber
\end{equation}
By integrating in the domain of the variable $\grs$ we get
\begin{equation} \label{orth1}
   (a^2-b^2) \int_0^{+\infty} \!\!\grs \barr{f_{k,b}} f_{k,a} d\grs = \left[ \grs \left(\barr{f_{k,b}} \frac{d}{d\grs}f_{k,a} - f_{k,a} \frac{d}{d\grs}\barr{f_{k,b}} \right)\right]_0^{+\infty}.
\end{equation}
From the right hand side we see that for orthogonality, when $a\neq \pm b$,  we need a function that itself and its derivative are finite at $\grs\rightarrow 0$, and at infinity they tend to zero faster than the overall $\grs$ goes to infinity.

If we consider $a, b >0$, such a function is the Hankel function of the second kind defined as
\begin{equation} \nonumber
  H^{(2)}_{\nu} (z) = J_\nu (z) - i Y_\nu (z),
\end{equation}
where $\nu, z$ are taken in such a manner so that we have
\begin{equation}\nonumber
  H^{(2)}_{i k} (-i a \grs) = J_{i k} (-i a \grs) - i Y_{i  k} (-i a \grs).
\end{equation}
The Bessel function of the first kind is given by the series
\begin{equation}
  J_\nu (z) = \left(\frac{1}{2} z\right)^\nu \sum_{j=0}^{\infty} (-1)^j \frac{\left(\frac{1}{4}z^2\right)^j}{j! \Gamma(\nu+j+1)},\nonumber
\end{equation}
while for that of the second kind it holds that
\begin{equation}
  Y_\nu (z) = \frac{\cos(\nu \pi) J_\nu(z)-J_{-\nu}(z)}{\sin(\nu \pi)}.\nonumber
\end{equation}
This means that at the limit $\grs\rightarrow 0$ the function $H^{(2)}_{i k} (-i a \grs)$ is oscillating but finite. It also holds that
\begin{equation} \label{rulder}
   \frac{d}{dz}\mathcal{C}_\nu (z) =\frac{\nu}{z} \mathcal{C}_{\nu} (z)-\mathcal{C}_{\nu+1} (z), \quad \nu, z \in \mathbb{C},
\end{equation}
where $\mathcal{C}_\nu (z)$ can be any of the $J_\nu (z)$, $Y_\nu (z)$, $H^{(1)}_\nu (z)$, $H^{(2)}_\nu (z)$ or their combinations. Hence we conclude that
\begin{equation}
  \lim_{\grs\rightarrow 0} \left[ \grs \left(\barr{f_{k,b}} \frac{d}{d\grs}f_{k,a} - f_{k,a} \frac{d}{d\grs}\barr{f_{k,b}} \right)\right] =0,\nonumber
\end{equation}
with $f_{k,a} =H^{(2)}_{i k} (-i a \grs)$. Note that $\barr{f_{k,b}} = \barr{H^{(2)}_{i k} (-i b \grs)}= H^{(1)}_{-i k} (i b \grs)$, where  $H^{(1)}_\nu(z) = J_\nu(z)+i Y_\nu (z)$ is the Hankel function of the first kind. According to \eqref{rulder} we obtain
\begin{subequations} \label{derivatives}
\begin{align}
  & \frac{d}{d \grs}f_{k,a}=\frac{d}{d \grs}H^{(2)}_{i k}(-i a \grs) = i a \left(H^{(2)}_{1+i k}(-i a \grs) + \frac{k}{a \grs} H^{(2)}_{i k}(-i a \grs)\right), \\
  & \frac{d}{d \grs}\barr{f_{k,b}}=\frac{d}{d \grs}H^{(1)}_{-i k}(i b \grs) = -i b \left(H^{(1)}_{1-i k}(i b \grs) + \frac{k}{b \grs} H^{(1)}_{-i k}(i b \grs)\right).
\end{align}
\end{subequations}
In order to calculate the limit of the right hand side of \eqref{orth1} at infinity we need only the leading terms as $\grs\rightarrow + \infty$. For the two Hankel functions involved, these are (remember that according to our assumption $a, b > 0$):
\begin{equation}
  H^{(2)}_{i k}(-i a \grs) = \frac{i \sqrt{2}}{\sqrt{\pi a  \grs}} e^{-\frac{\pi  k}{2}-a  \grs}, \quad H^{(1)}_{-i k}(i b \grs) = \frac{(1-i)}{\sqrt{i \pi b  \grs}}e^{-\frac{\pi  k}{2}-b  \grs}.\nonumber
\end{equation}
Substitution of the latter and of \eqref{derivatives} leads to
\begin{equation}
  \lim_{\grs\rightarrow +\infty} \left[ \grs \left(\barr{f_{k,b}} \frac{d}{d\grs}f_{k,a} - f_{k,a} \frac{d}{d\grs}\barr{f_{k,b}} \right)\right] = \lim_{\grs\rightarrow +\infty} \left[ e^{-\grs (a +b )}\left(-\frac{2 (a -b )}{\pi  \sqrt{a  b}}e^{-\pi  k} +\mathcal{O}(\grs^{-1})\right)\right] =0.\nonumber
\end{equation}
Thus, we can write \eqref{orth1} as
\begin{equation} \nonumber
   (a^2-b^2) \int_0^{+\infty} \!\!\grs \barr{f_{k,b}} f_{k,a} d\grs = -\frac{2 (a -b )}{\pi  \sqrt{a  b}}e^{-\pi  k} \lim_{\grs\rightarrow +\infty} \left( e^{-\grs (a +b )}\right) =0 , a\neq{b}.
\end{equation}
Hence, for $a\neq b$ the $f_{k,a}=H^{(2)}_{i k}(-i a \grs) $ is orthogonal (with weight $\grs$) to its conjugate possessing a different eigenvalue.

\subsection{Dynamical $n(t)$, eigenstate $\grc_{3}$}

The corresponding eigenstate is
\begin{equation}
  \psi_3 = e^{i p_3 \theta} \left(q_1 e^{i p_3 \sigma} + q_2 e^{-i p_3 \sigma} \right) .\nonumber
\end{equation}
Let us choose $q_2=-q_1$ so that
\begin{equation}
  \psi_3 = 2 q_1 i e^{i p_3 \theta} \sin(p_3 \sigma).\nonumber
\end{equation}
Note that it is important to take the sine function, a similar analysis does not work with the cosine.

To normalize the probability we need to take
\begin{equation}
  \begin{split}
    P &= \int\!\! \mu \psi_3(p_3')^* \psi_3(p_3) d\theta d\sigma= 4 q_1^2 \int_{-\infty}^{+\infty}e^{-i (p_3' -p_3) \theta}d\theta \int_{-\infty}^{\infty}\!\!e^{2\sigma} \sin(p_3'\sigma) \sin(p_3\sigma) d\sigma \\
    &=8\pi q_1^2 \delta(p_3'-p_3) \int_{-\infty}^{\infty}\!\!e^{2\sigma} \sin^2(p_3\sigma) d\sigma,\nonumber
  \end{split}
\end{equation}
where due to the presence of the delta we assume $p_3'=p_3$ in the remaining integral. Let us study the integral
\begin{equation} \label{Ilim}
  \begin{split}
    I=  &\int_{-\infty}^{\infty}\!\!e^{2\sigma} \sin^2(p_3\sigma) d\sigma = \lim_{L\rightarrow +\infty} \int_{-L}^{L}\!\!e^{2\sigma} \sin^2(p_3\sigma) d\sigma \\
     = &\lim_{L\rightarrow +\infty} \Big[\frac{e^{2 L}}{4(1+ p_3^2)}\left( p_3^2-p_3 \sin (2 p_3 L)-\cos (2 p_3 L)+1\right) \\
     & -\frac{e^{-2 L}}{4(1+ p_3^2)} \left(p_3^2+p_3 \sin (2 p_3 L)-\cos (2 p_3 L)+1\right)\Big].
  \end{split}
\end{equation}
We may now associate $p_3$ with the boundary and we consider it as a $p_3(L)$ such as the equation
\begin{equation} \label{alg1}
   p_3^2-p_3 \sin (2 p_3 L)-\cos (2 p_3 L)+1 = 4(1+ p_3^2) e^{-2 L},
\end{equation}
is satisfied. Note that the power in the exponent of the right hand side has to be exactly $-2L$ so that the limit appearing in \eqref{Ilim} will be neither zero, nor divergent. A multiplying constant could also be assumed at the right hand side, but it will get lost after the normalization so it is not really necessary .

\textbf{Lemma:} There exists a positive number $L=L_0$, so that the algebraic equation \eqref{alg1} has at least two real solutions with respect to $p_3$.

\textbf{Proof:} Consider the function
\begin{equation}
  f(p_3) = p_3^2-p_3 \sin (2 p_3 L)-\cos (2 p_3 L)+1 - 4(1+ p_3^2) e^{-2 L}.\nonumber
\end{equation}
First we observe that $f(0)=-4 e^{-2L}<0$. Then, we can see that for $p_3\geq 0$ we can write (by assigning the sine and cosine to their smallest possible values)
\begin{equation} \label{ineq1}
  f(p_3) > \left(1-4 e^{-2 L}\right)p_3^2 -p_3-4 e^{-2 L}.
\end{equation}
Now, assume $L=L_0$ such that $e^{2 L_0}>4$ then the equation $\left(1-4 e^{-2 L}\right)p_3^2 -p_3-4 e^{-2 L}=0$ has a positive root, say $p_3^{(+)}>0$. When $p_3>p_3^{(+)}$ the expression of the right hand side of the inequality \eqref{ineq1} and hence $f(p_3)$ is positive. As a result we have shown that for a finite $L=L_0$ we have a change in sign of $f(p_3)$ as $p_3$ ranges from zero to $p_3^{(+)}$. Hence, given the continuity of $f(p_3)$, a real solution $p_3=p_3^{(0)}$ to $f(p_3)=0$ exists in this region. Since $f(p_3)=f(-p_3)$, if $p_3=p_3^{(0)}$ is a solution, $p_3=-p_3^{(0)}$ is also a solution. Moreover,  we can see from the form of $p_3^{(+)}= \frac{1 + \sqrt{1+16 e^{-2 L_0}-64 e^{-4 L_0}}}{2 \left(1-4 e^{-2 L_0}\right)}$, that $p_3^{(+)}$ tends to zero the larger $L_0$ becomes. Thus the two roots $p_3=\pm p_3^{(0)}$ that are inside the region $(-p_3^{(+)},p_3^{(+)})$ tend to ``meet" at zero as $L\rightarrow \infty$.

The latter means $\lim_{L\rightarrow \infty} p_3^{(0)}(L) =0$, where $p_3^{(0)}(L)$ satisfies equation \eqref{alg1}. Then, this implies that $I=1$ and
\begin{equation}
  P = 8\pi q_1^2 \delta(p_3'-p_3),\nonumber
\end{equation}
so the normalization constant should be $q_1=\frac{1}{2(2\pi)^{1/2}}$ and the wave function becomes
\begin{equation}
  \psi_3 = \frac{i}{(2\pi)^{1/2}} e^{i p_3 \theta} \sin(p_3 \sigma),\nonumber
\end{equation}
with the values of $p_3$ concentrating strongly around zero for a space with an infinite boundary.
\end{appendices}

\section*{Acknowledgements}

\begin{figure}[h!]
\centering
  \begin{subfigure}[h]{0.265\linewidth}
    \includegraphics[width=\linewidth]{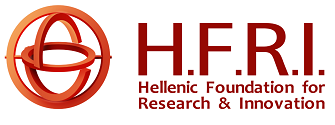}
  \end{subfigure}
  \begin{subfigure}[h]{0.2\linewidth}
    \includegraphics[width=\linewidth]{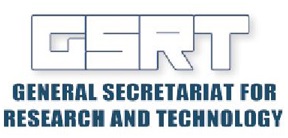}
  \end{subfigure}
\end{figure}
The research work was supported by the Hellenic Foundation for Research and Innovation (HFRI) and the General Secretariat for Research and Technology (GSRT), under the HFRI PhD Fellowship grant (GA.no.74136/2017).

\bibliography{bibliography}
\bibliographystyle{unsrt}

\end{document}